\documentstyle[epsf,preprint,aps]{revtex}
\tightenlines

\renewcommand{\theequation}{\arabic{section}.\arabic{equation}}

\newcommand{\asize}[1]{\renewcommand{\arraystretch}{#1}}

\newcommand{\be}{\begin{equation}}
\newcommand{\ee}{\end{equation}}
\newcommand{\ba}{\begin{array}}
\newcommand{\ea}{\end{array}}
\newcommand{\dst}{\displaystyle}
\newcommand{\tst}{\textstyle}

\newcommand{\bac}{\begin{array}{c}}
\newcommand{\bal}{\begin{array}{l}}
\newcommand{\baR}{\begin{array}{r}}
\newcommand{\bacc}{\begin{array}{cc}}
\newcommand{\ball}{\begin{array}{ll}}
\newcommand{\balr}{\begin{array}{lr}}
\newcommand{\barl}{\begin{array}{rl}}
\newcommand{\baccc}{\begin{array}{ccc}}
\newcommand{\barcl}{\begin{array}{rcl}}
\newcommand{\balcl}{\begin{array}{lcl}}
\newcommand{\barcll}{\begin{array}{rcll}}
\newcommand{\barll}{\begin{array}{rll}}
\newcommand{\barrclcl}{\begin{array}{rrclcl}}
\newcommand{\bacl}{\begin{array}{cl}}
\newcommand{\bacll}{\begin{array}{cll}}
\newcommand{\eac}{\end{array}}
\newcommand{\ber}{\begin{eqnarray}}
\newcommand{\eer}{\end{eqnarray}}

\newcommand{\ie}{{\it i.e.}}
\newcommand{\cf}{{\it cf.}}
\newcommand{\eg}{{\it e.g.\ }}

\newcommand{\tr}{{\rm tr}}

\newcommand{\Nbar}{{\overline{N}}}

\newcommand{\etabar}{{\overline{\eta}}}

\newcommand{\psibar}{{\overline{\psi}}}

\newcommand{\Gammabar}{{\overline{\Gamma}}}

\renewcommand{\d}{{{\partial}}}
\newcommand{\half}{\frac{1}{2}}

\newcommand{\Chat}{\hat{C}}
\newcommand{\dhat}{\hat{d}}

\newcommand{\ghat}{\hat{g}}
\newcommand{\khat}{\hat{k}}

\newcommand{\p}{^{\prime}}

\newcommand{\ra}{\rightarrow}

\newcommand{\mod}[1]{\quad (\mbox{mod}\;#1)}

\newcommand{\GG}{{\mathcal{G}}}

\newcommand{\II}{{\mathcal{I}}}

\newcommand{\LL}{{\mathcal{L}}}
\newcommand{\MM}{{\mathcal{M}}}
\newcommand{\NN}{{\mathcal{N}}}

\newcommand{\TT}{{\mathcal{T}}}
\newcommand{\VV}{{\mathcal{V}}}
\newcommand{\WW}{{\mathcal{W}}}

\newcommand{\sgn}{{\mathrm{sgn}}}
\newcommand{\sech}{{\mathrm{sech}}}

\newcommand{\arcsinh}{{\mathrm{arcsinh}}}

\renewcommand{\Re}{{{\mathcal{R}}e}}
\renewcommand{\Im}{{{\mathcal{I}}m}}

\newcommand{\txthalf}{{\textstyle \half}}

\newtheorem{theorem}{Theorem}

\begin{document}



\title{Single Spin Superconductivity: \\
Formulation and Ginzburg-Landau Theory}

\author{Robert E. Rudd$^{\dag}$ and Warren E. Pickett$^{\ddag}$}

\address{Naval Research Laboratory, Washington, DC 20375-5345}
\address{$^\dag$ SFA, Inc, 1401 McCormick Drive, Largo MD 20774}

\date{\today}
\maketitle

\tightenlines

\begin{abstract}
We describe a novel superconducting phase that arises due to a pairing
instability of the half-metallic antiferromagnetic (HM AFM) normal state.  This
single spin superconducting (SSS) phase contains broken time reversal
symmetry in addition to broken gauge
symmetry, the former due to the
underlying magnetic order in the normal state.  A classification of 
normal state symmetries leads to the conclusion that the HM AFM normal 
phase whose point group contains the inversion operator contains the least
symmetry possible which still allows
for a zero momentum pairing instability.
The Ginzburg-Landau free energy for the superconducting
order parameter is constructed consistent with the symmetry of the
normal phase, electromagnetic gauge invariance and the crystallographic
point group symmetry including inversion.
For cubic, hexagonal and tetragonal point groups, the possible symmetries of
the superconducting phase are classified, and the free energy is used
to construct a generalized phase diagram.  We identify the leading
candidate out of the possible SSS phases for each point group.
The symmetry of the superconducting phase is used to determine the cases
where 
the gap function has generic zeros (point or line nodes) on the Fermi surface.  
Such nodes always occur, hence thermodynamic properties will
have power-law behavior at low
temperature.  
\end{abstract}
\vskip 1cm
\footnoterule
$^{\dag}$ \parbox[t]{6in}{
{\small E-mail: ~~rudd@dave.nrl.navy.mil} }
\\
$^{\ddag}$ \parbox[t]{6in}{
{\small E-mail: ~~pickett@dave.nrl.navy.mil} } 


\newpage


\section{Introduction}

The pairing theory \cite{JRS,SU} of superconductivity and
superfluidity is based on a normal state with time 
reversal symmetry and inversion symmetry.  The former symmetry
requires that the two spin directions are related by symmetry, 
and spin-space rotations have played a central role in the classification 
of broken symmetry phases.  The latter symmetry is sufficient
to ensure that states with identical spin directions at 
$\vec k$ and $-\vec k$ are degenerate.
Electrons at $\vec k$ and $-\vec k$ can then be
paired to have zero total momentum and can be
classified by spin state as singlet or triplet.
The question of inversion symmetry and
time-reversal-breaking is central in the specification of the pairing states,
\cite{SU,VG1,UR,Blount,UR2,PWA,AB,OMO,GS}
and in this paper we present some new aspects of these relationships.

It has recently been pointed out \cite{WEP} that there is a
normal state with broken
time reversal symmetry that has a pairing instability in direct analogy
with that in BCS theory.\cite{JRS}  This magnetically 
ordered normal state, in which
{\em inequivalent} up-spin and down-spin magnetizations cancel
exactly, has been termed ``half-metallic antiferromagnet'' (HM AFM) by
van Leuken and de Groot.\cite{DG2}  This normal state, which we describe in 
Sec.\ \ref{sec-NormalState},
has considerable theoretical and probable technological interest in
itself.  Its lack of macroscopic magnetization means that
considerations of pairing do not have to confront the
question of competition between superconducting order and a pre-existing
magnetic field.

In this paper we outline in more detail the characteristics of this
``single spin superconductivity'' (SSS) phase.  In 
Sec.\ \ref{sec-NormalState} we review
the characteristics of the HM AFM state, which is the precursor
normal state of the SSS.  In Sec.\ \ref{sec-Pairing} we show that
the phenomenon maps onto the BCS model with minor but non-trivial
changes.  Comparison to liquid $^3$He, conventional BCS superconductors,
and more exotic heavy fermion superconductors in Sec.\ \ref{sec-Symmetry} 
demonstrates that the HM AFM state has the
minimum symmetry required of the normal state to allow $\vec q$ = 0
pairing instability, and that in SSS theory inversion symmetry plays a
role analogous to that of time-reversal symmetry in BCS theory.
In Sec.\ \ref{sec-GL} we provide the classification of all possible
order parameter symmetries for high symmetry crystal point groups, and 
present results of a symmetry analysis of the Ginzburg-Landau 
free energy that enumerates the possible SSS states for cubic,
tetragonal, and hexagonal crystals.

\section{The Normal State}
\label{sec-NormalState}
\setcounter{equation}{0}

\subsection{Half-metallic ferromagnetism}
\label{subsec-HM FM}

A half-metallic (HM) ferromagnetic (FM) electronic structure arises in
a ferromagnetic material when the Fermi level ($E_F$) of one spin
direction lies within a gap in the spectrum of the other spin direction.
\cite{DG1,IK}
The gap may occur in either the majority or the minority channel. 
In either case, we will take the up channel to be the metallic one.  
We also confine this discussion to stoichiometric compounds, which 
have an integer number of electrons per cell.  The system
of up spins then forms a metallic fermion liquid, while the down spins form an
insulating system that may be thought of as an inert background for the
purposes of studying low temperature, low energy processes.  This
specific occurrence, the placement of the Fermi level of the metallic
channel in a gap of the other channel, defines half-metallic 
character: the up-spin ``half'' of the electrons is metallic, while the
down-spin ``half'' is insulating.  Fig.\ \ref{fig-modDOS}(a) shows a model
spectrum of exchange split bands that leads to HM character.

Half-metallicity leads to several features 
of a crystalline solid that are qualitatively
distinct from conventional metallic ferromagnets.  Unlike in a
conventional FM, electron transport is 100\% polarized, and there are
no allowed low energy spin flips. 
In a common FM, the spin moment is a continuous quantity
whose value is determined by the balance of exchange energy and kinetic
energy. 
In a HM FM, however, the spin
moment is constrained to be precisely an integer ${\MM}$.  
This is so because the
insulating channel contains an integer number of filled bands, and
hence an integer spin ${\NN}_{\downarrow}$, the cell contains an
integer number of electrons ${\NN}_{tot}$, so the metallic up channel contains
an integer number of spins ${\NN}_{\uparrow}={\NN}_{tot}-
{\NN}_{\downarrow}$.  Then ${\MM}\equiv{\NN}_{\uparrow}-{\NN}_{\downarrow}$ 
is an integer, and the moment is ${\MM} \mu_B$ per cell.  
This follows for any placement 
of $E_F$ within the gap in the down-spin density of states.  The
application of a magnetic field H shifts up and downs spin bands by 
$\pm$g$\mu_B$H but does not change the occupation or the net spin moment.
Hence the spin susceptibility is precisely zero, which is a direct
consequence of the lack of
low-energy spin flips.  There is then no Stoner continuum
to damp spins waves by single spin flips of carriers.
In fact, the situation can be categorized as extreme spin-charge
separation in the carrier system, in which the spin degree of 
freedom has been separated from the charge fluctuations and frozen
out entirely.  
In this paper we do not address possible effects due to spin waves.

Perhaps the simplest example of a HM FM is CrO$_2$, in which the
moment is 2$\mu_B$.
\cite{KHS}  de Groot and collaborators\cite{DG1} have identified 
calculationally various Heusler alloys that are likely
HM FMs, and experimental work on several members (especially UNiSn and
NiMnSb) has been reported.  \cite{Fujii,hanssen}
Pickett and Singh\cite{WEPDJS} presented theoretical evidence that
the colossal magnetoresistance manganites, {\em viz.}
La$_{2/3}$Ca$_{1/3}$MnO$_3$, are HM in their low
temperature FM phase.  Recently several candidates for HM behavior
have been found with the double perovskite crystal structure.
\cite{WEP2}  These examples indicate that half-metallic character
is not a rare phenomenon.

\subsection{Half-metallic antiferromagnetism}
\label{subsec-HMAF}

\begin{figure}[tbp]
\epsfysize=15cm\centerline{\epsffile{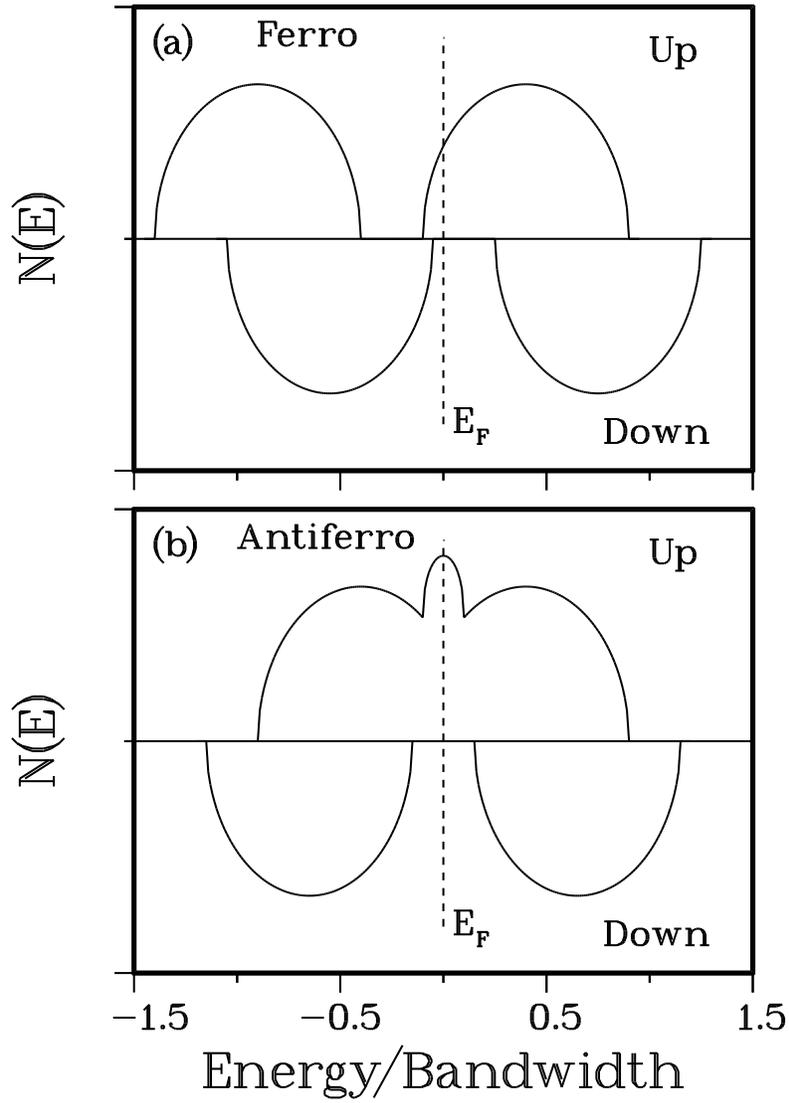}}
\caption{(a) Model spectrum with rigid exchange splitting that illustrates
a HM FM system.  (b) Model spectrum for a HM AFM illustrating that the 
channels must have different structure.  The peak at the Fermi level
is merely an artifact of the form and symmetry of the model.
\label{fig-modDOS}}
\end{figure}

It may occur that the integer spin moment ${\MM}$ in a half-metallic
system is zero.  This situation has been termed {\em half-metallic
antiferromagnetism} (HM AFM).\cite{DG2}  Its properties are like that of the
HM FM discussed above, with one essential difference: 
there is no
macroscopic magnetic field.  The HM AFM has 100\% polarized charge
transport without any net magnetization.  It must be kept in mind
that the HM AFM is not antiferromagnetic in the usual sense of the
term, as there is no symmetry operation that connects spin up
and spin down states or densities.  In fact, it is essential that
the two spins channels are electronically (and thus chemically) 
distinct, so a gap can occur in one channel only at the {\em same}
band filling. A model illustrating this situation is shown in 
Fig.\ \ref{fig-modDOS}(b).  This model uses the same two-band form of 
Fig.\ \ref{fig-modDOS}(a), but the bands for the two spin channels
must be displaced in opposite directions, reflecting the necessary 
inequivalence of the channels.

In a HM AFM the spins are precisely balanced, so there is no majority or
minority spin.  In this paper we will call (when the need arises)
the metallic channel
the ``up'' spin and the insulating channel the ``down'' spin.
For low energy and low temperature processes the insulating 
channel becomes inert and drops
out of consideration.  van Leuken and de Groot \cite{DG2} have suggested
one quintinary compound, intermetallic V$_7$MnFe$_8$Sb$_7$In in a 
Heusler-like crystal structure, that should be a HM AFM. 
One of the 
present authors\cite{WEP,WEP2} has found candidates for HM AFM states
within the class of magnetic double perovskites.  An example of a HM AFM
spectrum is shown in Fig. 1 for the double perovskite compound
La$_2$VCuO$_6$, calculated using accurate spin density functional
methods.\cite{WEPDJS}  This is nominally a Cu$^{2+}$, V$^{4+}$ compound for
which both ions have spin $\frac{1}{2}$.  When the spins are parallel
(top panel) the spectrum is that of a conventional metallic FM.  When the
spins are antialigned, however, the Fermi level falls where the spin up
density of states is large, but within a gap in the spin down channel.
This compound, and other double perovskites, are discussed in more 
detail elsewhere \cite{WEP2}.
 
We leave further discussion of proposed HM 
materials to future papers, and address below
the pairing instability of the HM AFM normal state, and its consequences.

\subsection{Experimental Consequences}

An important practical consideration is how a HM magnetic material
can be identified.  The anticipated properties (questions of many-body
corrections\cite{irkhin2} aside) have not previously been enumerated.
We include a partial list here to provide guidelines.  The general
feature of course that as the temperature is lowered through the
magnetic ordering (Curie or N\'eel) temperature $T_M$, the material 
changes from an nonmagnetic (conducting or non-conducting) system
to a metallic magnetically ordered system at low T where the spin excitations
are frozen out.\cite{wepdjs}  

\renewcommand{\theenumi}{\alph{enumi}}
\begin{enumerate}
\item {\em Metal with fully polarized transport at low T}. Metallic
resistivity, but vanishing magnetoresistance at low T.  
There is no clear signature in the
Hall or Seebeck coefficients.  At intermediate temperature there may be
a negative magnetoresistance for a HM FM, reflecting the field induced
increase in magnetic order and reduced spin scattering as the carriers
in one channel become non-conducting.  For HM AFM,
this field induced effect will not apply.
\item {\em Magnetic order}.  There is no obvious signature of the HM
character in the spin wave spectrum or temperature dependence of the
magnetic order parameter.
\item {\em Vanishing spin susceptibility}. However, core diamagnetism,
van Vleck (orbital) paramagnetism of the metallic channel, Landau diamagnetism
of the insulating channel, and temperature variation of the net order
of the local moments
will make the magnetic susceptibility difficult to analyze.  
%
\item {\em Non-Korringa behavior in NMR}.  This technique 
may provide the most direct
indication of HM character.  The longitudinal relaxation rate T$_1^{-1}$,
which is a measure of the conduction electron spin flips,
is  proportional to the product of the densities of states of each
spin channel [N$_{\uparrow}$(E$_F$)N$_{\downarrow}$(E$_F$)], which
vanishes for a HM phase.  The Knight shift, normally dominated by the
spin susceptibility in normal metals, should be small.
An NMR study of the proposed HM magnet UNiSn has been
reported.\cite{kojima}
\item {\em Spin-polarized electron spectroscopies}.  At first glance,
these spectroscopies seem ideal, but both
photoelectron emission and STM tunneling are sensitive to surface
properties.  In addition, the magnetic order may be different
at the surface, mitigating against HM character in the surface region.
Spin-polarized photoemission studies of CrO$_2$ were inconclusive.
\cite{cro2,KHS}
\item {\em Spin-polarized positron annihilation}.  This technique, which
takes advantage of the natural polarization of the positron beam, has
been claimed to establish within narrow bounds that NiMnSb is a
HM ferromagnet.\cite{hanssen}  It is a bulk probe.  
\item {\em Thermodynamic properties}.  We will show that SSSs 
necessarily have point or line nodes of the gap (not always the case in 
previously studied cases of triplet pairing).  The resulting gapless
excitations lead to heat capacity, penetration depth, thermal conductivity,
{\it etc.} that have power-law in T (or $\omega$) rather than
exponential.
\item {\em Tunnelling}.  Tunnelling between an SSS and a ferromagnet
will show a strong dependence on the direction of magnetization of
the ferromagnet.  Josephson coupling between an SSS and a 
singlet pairing superconductor should not occur. 
\end{enumerate}

\section{Pairing Instability in the HM AFM}
\label{sec-Pairing}
\setcounter{equation}{0}

The HM AFM is a single component fermion liquid as a result of underlying
magnetic order and electronic structure that renders one spin channel 
insulating. 
The lack of a microscopic magnetic field in a HM AFM suggests the
possibility that a
superconducting instability may occur in the metallic channel.  The Cooper
instability \cite{JRS} is spin-blind: the two fermions that 
undergo the pairing instability can have antiparallel spins as in BCS theory,
or they can have parallel spins as in $^3$He, 
and the instability is straightforwardly
extended to a spinless fermion system.  We now show that this
instability maps directly onto the BCS model of superconductivity\cite{JRS}
in a simple but not quite trivial manner.  This new superconducting state
has been called single spin superconductivity \cite{WEP}. 

\subsection{Formal Relationship to BCS Theory}
In BCS theory an electron in state $K$ is paired with its time-reversed
partner ${\TT}K$.  $K=(\vec k,\uparrow$) is an index that together with
its partner ${\TT}K=(-\vec k,\downarrow$) exhausts all states on the
($\uparrow$ and $\downarrow$) Fermi surface(s). 
In a SSS, an electron in state $K=(\vec k,+$) is paired with its inversion
partner ${\II}K=(-\vec k,-$).  To cover all states on the Fermi surface
once only, $K$ must range over only half of the Fermi surface, say the
`top' half with $k_z>0$ (hence the notation `$+$'), and states with
$k_z=0$ can be assigned to `$+$' and `$-$' components of the pairs also. 

In terms of the general two-body interaction
\be
 \hat {\VV} = \sum_{\vec k_1,\vec k_2,\vec k_3,\vec k_4}
   V_{\vec k_1,\vec k_2,\vec k_3,\vec k_4}
   a^{\dagger}_{\vec k_1}
  a^{\dagger}_{\vec k_2}a_{\vec k_3}a_{\vec k_4},
\ee
all terms except $\vec k_1=-\vec k_2\equiv\vec k,
\vec k_3=-\vec k_4\equiv\vec k\p$ are irrelevant for pairing.  
To count the pair states
properly, the full Brillouin sums in the above expression must be
expressed in terms of pair indices $K$.  
Using the anticommutation
relations of the electron field operators $a_{\vec k},a^{\dagger}_{\vec k}$,
and defining the pair annihilation operator $b_{\vec k}=a_{\vec k}
a_{-\vec k}$,
the resulting interaction is (making
the notational simplification $V_{\vec k,-\vec k,\vec k\p,-\vec k\p}
\rightarrow V_{\vec k,\vec k\p}$),
\be
  \hat {\VV}_{pair} =  \sum_{\vec k}^+ \sum_{\vec k\p}^+ \,
        \left[ V_{\vec{k},\vec{k}\p} +
         V_{-\vec k,-\vec{k}\p}
       - V_{\vec{k},-\vec{k}\p} - V_{-\vec{k},\vec{k}\p} \right]
  b^{\dagger}_{\vec k}b_{\vec k\p}
 \equiv \sum_{\vec k}^+ \sum_{\vec k\p}^+ \, U_{\vec{k},\vec{k}\p}
b^{\dagger}_{\vec k}b_{\vec k\p}.
\label{symmInt}
\ee
The `+' sign indicates that the summation extends only over the `top'
half $k_z>0$ of the Brillouin zone.
The symmetry of the matrix element noted by Sigrist and Ueda\cite{SU}
results in all terms in the braces being identical, so
$U_{\vec k,\vec k\p}=4 V_{\vec k,\vec k\p}$.  We return to the
implications of the form of the matrix element in the next subsection.
From $V_{\vec{k},\vec{k}\p}
\equiv \langle \vec{k}\p, -\vec{k}\p | V | \vec{k}, -\vec{k} \rangle
= \langle K\p,{\II}K\p | V | K,{\II}K \rangle
= V_{K,K\p}$, and similarly for $U_{K,K\p}$, and the expression 
for the kinetic energy in terms of
pair labels (running over half of the Fermi surface),
the pairing Hamiltonian then is
\be
H_{pair}= \sum_K \epsilon_{\vec k} (a^{\dagger}_K a_K + 
a^{\dagger}_{{\II}K} a_{{\II}K})
 + \sum_K \sum_{K\p} U_{K,K\p} b^{\dagger}_K b_{K\p}
\ee
where $b_K \equiv a_K a_{{\II}K}$.
Note that we do not use the BCS convention of separating out a negative sign
from the interaction matrix elements. 

The Bogoliubov-Valatin transformation \cite{BV}
is analogous to its form in BCS theory,
\begin{eqnarray}
a_K & = & u_K \alpha_K + v_K \alpha^{\dagger}_{{\II}K}, \\
a^{\dagger}_{{\II}K} & = & u^*_{{\II}K} \alpha^{\dagger}_{{\II}K}
                 + v^*_{{\II}K} \alpha_K,
\end{eqnarray}
and the formalism of SSS pairing theory maps onto BCS theory.
Specifically, the SSS ground state is
\be
\Phi _0 =
\prod _K (u_K + v_K b_K^{\dagger})
\Phi _{vac}
\ee
and the gap function is given by
\be
\Delta_K =\sum_{K\p}
U_{K,K\p}
\langle b_K \rangle .
\label{gapFunc}
\ee
The gap is a scalar, {\it i.e.} it has no spinor indices, which distinguishes
SSS from all superconducting systems discussed previously.
The gap equation is formally identical to BCS:
\be
\Delta_K= -\sum_{K\p} \frac {U_{K,K\p}}
 {2E_{K\p}} \Delta_{K\p}
\tanh \left( \txthalf \beta E_{K\p} \right)
\label{GapEqnK}
\ee
where $\beta = 1/kT$ is the inverse temperature.
With $\mu$ denoting the chemical potential,
\be
E_K \equiv E_{\vec k} = [(\epsilon_{\vec k} - \mu)^2 +
|\Delta_{\vec k}|^2]^{\frac{1}{2}} 
\label{quasiSpec}
\ee
is the quasiparticle excitation energy, which is even in $\vec k$.

It is useful to express the gap equation as usual in terms of a full
Brillouin zone summation.  It is readily verified that extending
the sum over the full zone introduces an expected factor of $\frac{1}{2}$,
and the gap equation,
\be
\Delta_{\vec{k}} = -\sum_{\vec{k}\p} \frac {W_{\vec{k},\vec{k}\p}}
 {2E_{\vec{k}\p}} \Delta_{\vec{k}\p}
\tanh \left( \txthalf \beta E_{\vec{k}\p} \right)
\label{GapEqn}
\ee
where $W_{\vec k,\vec k\p}=\frac{1}{2}U_{\vec k,\vec k\p}=
2 V_{\vec k,\vec k\p}$, indicates
the final formal equivalence to the BCS equation.

\subsection{Simple Consequences of Single Spin Pairing}
The combination of matrix elements in Eq.\ (\ref{symmInt}), and the 
symmetry noted by Sigrist and Ueda,\cite{SU}
\be
V_{\vec k,\vec k\p}=- V_{-\vec{k},\vec{k}\p} = 
 - V_{\vec{k},-\vec{k}\p} = V_{-\vec k,-\vec k\p},
\ee
indicate explicitly the $k$-space structure 
that is necessary for SSS pairing.
A $\vec k$-independent
attractive potential $V_{\vec k,\vec k'}=-\bar V$, which leads to
singlet pairing in the BCS model, contributes nothing toward SSS
pairing; likewise, a $\vec k$-independent repulsion is harmless.
The simplest form of such coupling, which is odd in 
both $\vec k$ and $\vec k\p$,
is of the form ${\WW} \hat k \cdot \hat k\p$.  When this 
is substituted into the gap equation Eq.\ (\ref{GapEqn}) (see below), 
a non-vanishing
solution requires ${\WW} < 0$.  Thus the pairing interaction must be
attractive for small angle (``forward'') pair scattering 
and repulsive for large angle scattering of pairs.
This behavior is reminiscent of the situation in high $T_c$
theory, where the spin-fluctuation picture has an everywhere-positive
interaction, which peaks at large $q$ (more specifically, at
$(\pi/a,\pi/a))$.\cite{Scalapino}
That type of interaction favors a $d_{x^2-y^2}$ symmetry of $\Delta$
for singlet pairing.

Using the pairing interaction in normalized form
\be
  W_{\vec k,\vec k\p} = -|{\WW}|\frac{\vec k \cdot \vec k\p}{k_F^2/3} ,
\label{kxkp}
\ee
where $k_F$ is the Fermi wave vector,
the T=0 gap equation can be solved readily for several trial gap symmetries.
Gap functions of the form $\Delta_{\vec k}\propto \hat d \cdot \hat k$,
for some constant vector $\hat d$, give the lowest order possibilities
(in terms of polynomials of the components of $\vec k$).  If $\hat d$ is
real, or purely imaginary, its direction can be taken as the $\hat z$
axis ($\hat d$=(0,0,1)) so $\Delta_{\vec k} \propto k_z$ with a line of
nodes on the equator of the Fermi surface.  Complex $\hat d$ can be
represented by $\hat d$=(1,$i$,0)/$\sqrt 2$, in which case $\Delta_{\vec k}
\propto k_x + i k_y$ with point nodes at the poles.  We also consider
the ``highest symmetry'' single dimensional function $\Gamma_1^{(-)}
\propto k_x k_y k_x (k_x^2-k_y^2)(k_y^2-k_z^2)(k_z^2-k_x^2)$
(see the following Section) as an exotic possibility -- it has nine
lines of nodes.

We treat the
usual weak-coupling case, where
\begin{equation}
\sum_{\vec k}\rightarrow \int_{-\omega_c}^{\omega_c}d\epsilon N(\epsilon)
 \int \frac {d\Omega(\hat k)}{4\pi} \rightarrow
 N(0)\int_{-\omega_c}^{\omega_c}d\epsilon \int \frac{d\Omega({\hat k})}{4\pi},
\end{equation}
where $N(\epsilon)$ is the density of states per spin which is assumed
to be constant over the energy scale $\omega_c$
of the pairing interaction.
We display in Fig. 3 the resulting T=0 gap value, relative to the energy
cutoff $\omega_c$, versus the coupling strength $\lambda=N(0)|{\WW}|$
for these gap symmetries.
The BCS result is given for comparison.
Given the same coupling $\lambda$, it is
evident that the zero temperature gap magnitude
is comparable to the BCS value, even for the $\Gamma_1^{(-)}$ function.

It is straightforward to obtain the limiting behaviors of the zero
temperature gap from Eq.\ (\ref{GapEqn}).  In the weak-coupling limit,
\begin{equation}
\Delta_{rms}=2\omega_c e^{-1/\lambda} C_1(1+C_2 e^{-2/\lambda}),
\end{equation}
while for large $\lambda$ the asymptotic form is
\begin{equation}
\Delta_{rms}=\omega_c \lambda (D_1 - \frac{D_2}{\lambda^2}).
\end{equation}
Here $C_1, C_2, D_1, D_2$ are symmetry dependent
constants (\cf ~Appendix B).  
This latter relation explains the linear behavior at larger
$\lambda$ that is evident in Fig. 3.  Note that strong coupling corrections
will change this to a $\sqrt \lambda$ behavior.\cite{allendynes}

At $T=T_c$, $E_{\vec k}\rightarrow \epsilon_{\vec k}$ (we take $\mu=0$).
Taking $\vec k$ along $\hat d$
(or along one non-zero component if $\hat d$ is complex), the only
difference from the corresponding BCS equation is an angular
integral.  This integral is unity, however, due to the expansion of
$W_{\vec k, \vec k\p}$ in normalized functions.  Thus the equation for
$T_c$ versus coupling constant $\lambda=N(0)|{\WW}|$
is identical in form to that of BCS:
\begin{equation}
\frac {1}{\lambda}=\int_0^{\omega_c} d\epsilon
 \frac{\tanh (\beta_c \epsilon /2)}{\epsilon},
\label{eq:tc}
\end{equation}
where $\beta_c=1/k_B T_c$.

Although there is every reason to expect that SSS will arise in the
appropriate transition metal (or $f$ electron) compounds, there is not
yet any expectation of high $T_c$.  For one thing, the necessary 
interaction $W_{k,k\p}$ is of a particular kind (see above); however,
this is also the case for the $d$ wave scenario in high $T_c$
cuprates.  More to the point, however, is that at larger $T$
transverse spin fluctuations increase strongly and tend to reduce the
AFM order parameter, finally causing the system to revert to 
the paramagnetic state above the N\'eel
temperature.  The manner in which the HM AFM state will thereby be
weakened has not yet
been explored.

\section{Symmetry: Relations to Previous Theory}
\label{sec-Symmetry}
\setcounter{equation}{0}

It is instructive to clarify the relationship between the degree of symmetry of
the normal state and the degree of richness of broken symmetry in the
condensed phase.  $^3$He has the highest symmetry possible in its (liquid)
normal state.  It has continuous real space ${\LL}$ and spin rotation
${\cal S}$ symmetries, it has time reversal ${\TT}$ and inversion ${\II}$, and
of course gauge symmetry U(1).  The group of its normal phase then is
${\LL}\times{\cal S}\times{\TT}\times{\II}\times$U(1).  In the condensed
superfluid phase, U(1) is a broken symmetry and one or more of the
other symmetries can be broken concomitantly.  Much work has been done
to characterize the more likely cases among the infinite possibilities
(infinite because the relative angular momentum L of the pair can be
any non-negative integer).  The observed phases correspond to 
particular states within the (pair spin and orbital angular momentum
quantum numbers) $S=1$, $L=1$ complex that is described by an
18 component order parameter in general.

When considering pairing in a crystal, the continuous real space rotation
symmetry is replaced by the finite crystalline point group G (see 
Table \ref{table-sym},
where the classifications of this subsection are collected).  The 
group of the normal state then is 
G$\times{\cal S}\times{\TT}\times{\II}\times$U(1).  The number
of broken states is finite because the space of basis functions of irreps is
spanned by a few small-L sets.  The necessary values of L for cubic crystals
are presented under the BCS case in Table \ref{table-sym}.  
The allowed symmetries in
cubic, tetragonal, and hexagonal crystals have been exhaustively categorized.
\cite{SU,VG}

In crystals with strong spin-orbit coupling, it may be appropriate to
consider the spin as frozen into the crystalline lattice, in which case
spin rotation is no longer a separate symmetry.  (On this matter there
are arguments both {\em pro} and {\em con} in the literature.)  Then
the normal state symmetry is lowered further, and the number of
distinct broken symmetry states is further reduced.
The next lowest symmetry situation, where inversion symmetry is absent from
the point group, has been considered by Polu\'{e}ktov \cite{pol} with the
spins frozen to the lattice, distinction between singlet and triplet 
pairing vanishes and the order parameter becomes a low symmetry combination.

For the SSS discussed in this paper, the two spin systems are 
inequivalent, so time-reversal and spin rotation symmetries are strongly broken
by the normal state.  The symmetry of the normal state is described
by G$\times{\II}\times$U(1).  The number of distinct possibilities
for broken symmetry states, which we enumerate below, is reduced
still further.  In fact, this state of affairs gives
the lowest possible symmetry normal state that still allows pairing
in the usual sense of zero net momentum $\vec Q$ of the 
pair.  Inversion symmetry ensures that
$\epsilon_{\vec k} \equiv \epsilon_K = \epsilon_{{\II}K} 
\equiv \epsilon_{-\vec k}$, so that
if $\vec k$ lies on the Fermi surface then so does $-\vec k$, and these two
states can pair to total momentum $\vec Q = \vec k + (-\vec k) = 0$.  Without
inversion symmetry, \ie for G$\times$U(1) normal state
symmetry, $\vec Q=0$ pairing is not allowed.  This result indicates
that inversion, not time-reversal symmetry, is the minimal symmetry
requirement for a pairing instability. For convenience, 
inversion $\II$ will be considered part of the point group $G$ below.

Because for a SSS state time-reversal symmetry is already broken in
the normal state, it is natural to expect (and indeed we find) that
broken symmetry phases with unusual properties are likely to arise.
Such considerations occupy the rest of the work reported here. 


\section{Ginzburg-Landau Free Energy}
\label{sec-GL}
\setcounter{equation}{0}

\subsection{Allowed Superconducting Phase Symmetries}
\label{subsec-SCsym}

The spontaneous symmetry breaking at the superconducting phase transition is
governed by the free energy.  Below $T_c$ the minimum
is a superconducting phase whose order parameter breaks the $U(1)$ gauge
invariance and possibly other symmetries as well.  
This physics may be described through a 
phenomenological Ginzburg-Landau free energy that describes the mean
field theory of the superconductor in terms of a few parameters 
related to matrix elements of the effective potential.  Despite 
its simplicity, the mean field free energy captures
all of the generic information about the allowed symmetries of the
superconducting phase.

The order parameter describing the Cooper pair condensate is taken
to be the gap function, 
\be
\Delta_{\vec{k}} =\sum_{\vec{k}^{\prime}}
W_{\vec{k},\vec{k}^{\prime}}
\langle a_{\vec{k}^{\prime}} a_{-\vec{k}^{\prime}} \rangle ,
\ee
which was introduced in Equation (\ref{gapFunc}).
%
%
Below $T_c$ the gap function is non-zero, and it transforms 
under the full symmetry group of the normal phase.  
In particular, it is not invariant under 
the $U(1)$ Abelian gauge symmetry (and the concomitant global $U(1)$ 
of electron number),  
since Cooper pairs have electric charge $-2$ (electron number $2$).
The $U(1)$ symmetry is broken to the cyclic group ${\mathbf{Z}}_2$.
The gap function may break other symmetries as well, and this is the
main focus of this Section.   


The gap function must form an irreducible representation (irrep) of the
symmetry group, $\GG = G \times U(1)$, at least in the
vicinity of the phase transition.  This follows from the fact that 
near the transition, the gap function satisfies the linearized gap equation,
which for a spherical Fermi surface is given by
\be
\barcl
{\dst \Delta_{\hat{k}} } & = & { \dst -
\gamma \left(\omega_c/kT_c \right) 
\int \! d\hat{k}\p \, W_{\hat{k},\hat{k}\p} 
\Delta_{\hat{k}\p} } \\ & & {\dst 
{\mathrm{with}}~~\gamma (x) = N(0) \int _{-x}^x dy \,
\frac{\tanh \left( \txthalf y \right) }{y} }
\eac
\ee
where $\omega _c$ is the cutoff for the interaction. 
This is a $\GG$-invariant eigenvalue equation where $\Delta$ is an 
eigenvector and therefore must transform as a member of an irrep of $\GG$.
As usual, we start with the assumption that the transition 
temperature for the first irrep to condense is much higher than that of the
others.  This lets us focus on each irrep separately.

Under a U(1) gauge transformation $a_{\vec k} \ra e^{-i\varphi}a_{\vec k}$,
$\Delta \ra e^{-2i\varphi} \Delta$.  
$\Delta$ may transform under the point
group, $G$, as well.  Consider the case where there is an
element of $G$ that sends $\Delta$ to another function which is not
related to $\Delta$ by a gauge transformation; \ie ~there is 
$g\in G$ such that $\ghat \Delta = \Delta ^\prime$ where
$|\Delta | \ne |\Delta ^\prime|$.  This breaks the point group symmetry 
to a subgroup $H$ of $G$ that does leave 
$\Delta$ invariant up to a gauge transformation.  This non-trivial 
form of spontaneous symmetry 
breaking occurs exactly when $\Delta$ is in an irrep $\Gamma$ of 
$G$ whose dimension is greater than one \cite{foot-equivar}.

The overall symmetry breaking scheme may be described as follows.  When 
$\Delta$ is in the irrep $\Gamma$ of $G$, there is a maximal subgroup $H$
of $G$ under which $\Delta$ transforms as a one dimensional irrep
$\Gamma ^{\prime}$.  Then $\GG$ is broken to the little group $H$,
$\GG \ra H$.  All such
decompositions of irreps of the crystallographic points groups
are tabulated in ``compatibility tables'' 
\cite{foot-CG}.  
It is merely a matter of looking up the maximal 
subgroup $H$ in which each distinct one dimensional irrep
appears.
This classifies all of the possible superconducting phases 
according to symmetry.

There is one subtlety in this analysis.  The gap function is not gauge
invariant, so it is not a physical observable.  The physical residual
symmetry group $H_{phys}$ may be larger than $H$.  
For example, the spectrum of quasiparticle excitations (\ref{quasiSpec})
has the symmetry of the gauge invariant product 
$\Delta ^{*} \Delta$.  If $\Delta$ is in a complex irrep, 
there may exist an element $g$ of $\GG$ that switches $\Gamma ^{\prime}$ and 
its complex conjugate $\Gammabar ^{\prime}$;
that is, $\ghat \psibar = \psi$ for $\psi \in \Gamma ^{\prime}$.  
Then $g$ leaves $|\Delta |^2$ invariant, but it is not an
element of $H$.  The residual symmetry of the physical
observables below $T_c$ is the group $H_{phys}$ generated
by $H$ and $g$ ($H_{phys} \cong H \times \mathbf{Z}_2$).  
On the other hand, if there is no such $g$ such as when the irrep 
is real, then $H_{phys} = H$. 

It is convenient to expand the gap function in terms of
explicit representatives of the irrep $\Gamma$.  A 
general form of the mode expansion of $\Delta _{\vec k}$ is
\be
\Delta _{\vec k} = 
\sum _{\Gamma ,m} \sum _{n_i} \eta _m(\Gamma ; n_i ) 
\, \psi ( \Gamma, m; n_i; \vec{k} )
\label{gapModes}
\ee
where $\eta _m(\Gamma ; n_i )$ is a complex coefficient, 
$m=1,\cdots ,\dim \Gamma$ labels the components of $\Gamma$
and the indices $n_i$ distinguish polynomials of different degrees.
(For multiple or aspherical Fermi surfaces Allen's Fermi surface 
harmonics \cite{pba}
would be used.)
A suitable choice of basis elements $ \psi ( \Gamma, m; n_i=0; \vec{k} )$
is given in Table \ref{table-basis}.  A complete basis of higher modes 
of the linearized gap equation may be constructed using standard
techniques.

\subsection{Construction of the Free Energy}
\label{subsec-Fconstruct}

At this point we have identified the possible residual symmetries of the
superconducting phase.  It remains to show how each is realized as
a minimum of the free energy.  The Ginzburg-Landau free energy,
$F$, is a functional of the mean field $\Delta _{\vec k}$ 
in which the fluctuations about the mean field have been integrated 
out \cite{Wilson,Shankar}.
It must be invariant under the full symmetry group $\GG$ which
describes the physics of the normal state.  
This constrains the combinations of the modes of $\Delta _{\vec{k}}$
(\ref{gapModes}) that enter $F$.  
Only invariant ($\Gamma ^+_1$) combinations contribute.
In a perturbative expansion of $F$ in terms of 
$\Delta _{\vec k}$, the form of the low order terms is highly
constrained by $\GG$-invariance.

The group symmetry imposes a very restricted form.  
Many terms that could appear in the free energy vanish.
This is usually computed with a Clebsch-Gordon decomposition 
\cite{SU,foot-CG}, but we develop a more powerful technique 
in which the group symmetry is used directly. 
The result is an explicit computation of the allowed terms in the 
free energy to arbitrarily high order in perturbation theory.

The group symmetry imposes a number of constraints.
Gauge invariance requires that the polynomial
have equal numbers of $\eta$'s and $\etabar$'s.
The restrictions due to the point group
are implemented as follows.  Each group
operation may be considered to act on the $\eta$'s in a way that leaves
the free energy invariant.  If $\Delta$ is transformed under an 
operation from $\GG$, it may be restored to its original form by a linear
transformation of the constants $\eta _m$; that is, they form the
contragredient representation of $\Gamma$ of $G$ with
charge 2 under global $U(1)$.  
The free energy is invariant under 
$\GG$, so it must be an invariant polynomial in $\eta _m$ 
under the action of $\GG$.
In addition to the symmetry constraints, 
the free energy must be real and bounded below for stability.

Invariant polynomials have been studied extensively in the mathematics
literature.
In particular, invariant polynomials for the symmetric and alternating 
groups \cite{FH} 
and for Abelian and non-Abelian gauge 
groups \cite{Zelobenko} have been constructed explicitly.  
These polynomials play an important role in gauge theory \cite{AS,Fpow}.
Invariant polynomials for $G\times U(1)$ require an extension of this
theory, and since it has not been discussed in the condensed matter literature,
we will give some details.


Since $G$ is a finite group, its action on the coefficients $\eta _j$
is isomorphic to a direct product of finite simple groups--in particular
the symmetric groups, $S_3$ and $S_2$, the alternating groups $A_3$ and
$A_2$ and the cyclic groups, ${\mathbf{Z}}_3$ and ${\mathbf{Z}}_2$.
The form of invariant polynomials for each of these groups is well 
known, and our task is to form combinations of them that are 
invariant under $G\times U(1)$.

The free energy is almost trivially constructed for the 
one dimensional representations.  Gauge invariance
requires that the free energy be a function of 
$|\eta _1|^2$.  This is also invariant under the point
group operations.  The perturbative expansion of the
free energy takes the form
\be
F = \alpha \, |\eta _1|^2 +
\beta \, |\eta _1|^4 + \gamma \, |\eta _1|^6
 + \cdots ~~~~~~~~~~~\mathrm{for}~\dim \Gamma = 1
\label{FoneDim}
\ee
where $\alpha, \beta$ and $\gamma$ are parameters describing the
expansion of the effective potential for the order parameter in terms of
the basis elements of the irreducible representation.  
Note that the $\vec{k}$ integrals for matrix elements of the
effective potential $W_{\vec{k},\vec{k}\p}$ have been included in these 
parameters, so they encode the physics. 
In general,
there are exponentially small corrections to this perturbation
series of the form $P(|\eta _1|^2) \, e^{-1/(\alpha \p |\eta _1 |^2)}$
where $P$ is a polynomial.  These non-perturbative corrections can be
important at very low temperatures or in strongly coupled systems,
but they are beyond the scope of this paper.

The treatment of higher dimensional representations is more involved.
The first step is to find a basis for the point group irrep that 
respects the simple group decomposition.  Each simple group generator
should correspond to a specific element of $G$.  This makes the
symmetries of the invariant polynomials manifest, permitting a
straightforward description.  Using invariant polynomial techniques,
we will construct the free energy for $\Gamma ^-_3$ and 
$\Gamma ^-_{4,5}$ of $O_h$, 
$\Gamma ^-_{5,6}$ of $D_{6h}$ and
$\Gamma ^-_{5}$ of $D_{4h}$. 

Consider the two dimensional representation 
$\Gamma _3^-$ of $O_h$.
The standard real basis shown in Table \ref{table-basis} 
does not respect the simple group decomposition.  
The appropriate basis is composed of the complex functions
\be
\barcl
{\dst
\psi ^{\prime}_1 
} & = & {\dst
 [\psi _1 (\Gamma _3^-) + 
i \, \psi _2(\Gamma _3^-)]/\sqrt{2} ~ = ~
k_xk_yk_z ( k_z^2 + \omega k_x^2 + \omega ^2 k_y^2 )
} \\ {\dst
\psi ^{\prime}_2 
} & = & {\dst 
[ \psi _1 (\Gamma _3^-) - 
i \, \psi _2(\Gamma _3^-)]/\sqrt{2} ~ = ~
k_xk_yk_z (k_z^2 + \omega  ^2 k_x^2 + \omega k_y^2) .}
\eac
\ee
The action of $O_h$ on this basis is isomorphic to 
${\mathbf{Z}_3} \times {\mathbf{Z}}_2 \times \II$, where 
${\mathbf{Z}_3} = \{ E, C_{3\alpha}, C_{3\alpha}^{-1}\}$, 
${\mathbf{Z}_2} = \{ E, C_{2a}\p \}$ and $\II$ is inversion
(another ${\mathbf{Z}_2}$).
In addition to $U(1)$, we only need to consider the following two 
point group operations:
\be
\Chat _{3\alpha} = \left( 
\bacc
\omega ^2 & 0 \\
0 & \omega 
\eac
\right), \hspace{1cm}
\Chat _{2a}\p = \left( 
\bacc
0 & -1 \\
-1 & 0  
\eac
\right) .
\ee
Consider a monomial 
$(\eta ^{\prime}_1)^{N_1}\, (\eta ^{\prime}_2)^{N_2}\, 
(\etabar ^{\prime}_1)^{\Nbar _1}\, (\etabar ^{\prime}_2)^{\Nbar _2}$
in the perturbative expansion of the free energy.
Invariance under $O_h \times U(1)$ imposes the following constraints 
\begin{enumerate}
\item $U(1)$: $N_1 + N_2 = \Nbar _1 + \Nbar _2$.
\item $C_{3\alpha}$: $N_1 - N_2 - \Nbar _1 + \Nbar _2 \equiv 0 \mod{3}$.
\item $C_{2a}\p$: invariance under $\eta ^{\prime}_1 \ra - \eta ^{\prime}_2,   
  \eta ^{\prime}_2 \ra - \eta ^{\prime}_1$.  
\end{enumerate}
This leads to a basis of invariant polynomials of the form
\be
\barcl
 &  & {\dst
\left( |\eta ^{\prime}_1|^{2\beta}\, |\eta ^{\prime}_2|^{2\gamma }\, 
 + |\eta ^{\prime}_1|^{2\gamma}\, |\eta ^{\prime}_2|^{2\beta } \right) \, 
\Re \left[ (\eta ^{\prime}_1 \etabar ^{\prime}_2)^{3\alpha } \right] }
\\
{\dst {\mathrm{and}~~~~}
} & ~ & {\dst
\left( |\eta ^{\prime}_1|^{2\beta}\, |\eta ^{\prime}_2|^{2\gamma }\, 
 - |\eta ^{\prime}_1|^{2\gamma}\, |\eta ^{\prime}_2|^{2\beta } \right) \, 
\Im \left[ (\eta ^{\prime}_1 \etabar ^{\prime}_2)^{3\alpha } \right] }.
\eac
\ee
They can be reexpressed in terms of the polynomial generators for 
$\Gamma _3^-$ of $O_h$
\be
\barcl
{\dst P_1} & = & 
{\dst |\eta ^{\prime}_1|^{2} + |\eta ^{\prime}_2|^{2} } \\
{\dst P_2} & = & 
{\dst 4 |\eta ^{\prime}_1\eta ^{\prime}_2|^{2} } \\
{\dst P_3} & = & 
{\dst 8 \Re \left[ (\eta ^{\prime}_1 \etabar ^{\prime}_2)^{3} \right] } \\
{\dst P_4} & = & 
{\dst 8 \left( |\eta ^{\prime}_1|^{2} - |\eta ^{\prime}_2|^{2} \right) ~
\Im \left[ (\eta ^{\prime}_1 \etabar ^{\prime}_2)^{3} \right] } \\
\eac
\label{genG3Oh}
\ee
where the complete set of invariant basis elements are powers of these
four elements, in one of the two forms
\be
\barcl
{\dst 
P_{(m,n,p)}^+  } & = & {\dst
P_1^{n}P_2^{p}P_3^{m} } \\
{\dst P_{(m,n,p)}^-  } & = & {\dst
P_1^{n}P_2^{p}P_3^{m}P_4. }
\eac
\label{genBasisG3Oh}
\ee
These invariant polynomials have not been constructed previously.

Using the invariant basis the most general form of the perturbative
expansion of the free energy may be written
\be
\barcl
{\dst F(O_h(\Gamma _3^-))} & = & {\dst \sum _{m,n,p} 
 F_{(m,n,p)}^+ \, P_{(m ,n ,p)}^+ 
(\eta ^{\prime}_1, \eta ^{\prime}_2)
 + F_{(m,n,p)}^- \, P_{(m ,n ,p)}^- 
(\eta ^{\prime}_1, \eta ^{\prime}_2)
} \\ & = & {\dst
\alpha \, (|\eta ^{\prime}_1 |^2 + |\eta ^{\prime}_2 |^2) +
\beta _1\p \, (|\eta ^{\prime}_1 |^2 + |\eta ^{\prime}_2 |^2)^2 +
4\beta _2\p \, |\eta ^{\prime}_1 \eta ^{\prime}_2 |^2 + \cdots }
\eac
\label{FG3Oh}
\ee
where $\alpha = F^+_{(0,1,0)}$, $\beta _1\p = F^+_{(0,2,0)}$, 
$\beta _2\p = F^+_{(0,0,1)}$, etc.  
For comparison, 
Sigrist and Ueda \cite{SU} use the coefficients 
$\beta _1 = \beta _1\p + \beta _2\p$ and
$\beta _2 = \beta _2\p$.
The minima of this free energy will be studied in the next Subsection. 

The next case is the two dimensional representations $\Gamma _5^-$ 
and $\Gamma _6^-$ of $D_{6h}$.  
The analysis is essentially identical to the case of $\Gamma _3^-$
of $O_h$.  
Again the action of $D_{6h}$ on these two irreps is isomorphic to
${\mathbf{Z}_3} \times {\mathbf{Z}}_2 \times \II$, where 
${\mathbf{Z}_3} = \{ E, C_{3z}, C_{3z}^{-1}\}$, and
${\mathbf{Z}_2} = \{ E, C_{2x} \}$. 
The difference in the two cases comes from the way 
${\mathbf{Z}}_2 \times \II$ is embedded in the group; \eg
$\Chat _{2y} (\Gamma _5^-) = I \Chat _{2y} (\Gamma _6^-)$.
A complex basis is necessary to make the simple group
decomposition manifest
\be
\barcll
{\dst
\psi ^{\prime}_1 
} & = & {\dst
[\psi _1  + i \, \psi _2]/\sqrt{2} } \\
& = & {\dst
(k_x + ik_y)/\sqrt{2}
} & {\dst \mathrm{for}~~\Gamma _5^- }\\
& = & {\dst
k_yk_z (k_x + ik_y)( k_y^2 - 3k_x^2 )/\sqrt{2}
} ~~~~~~~~~~~ & {\dst \mathrm{for}~~\Gamma _6^- } \\ [2mm] {\dst
\psi ^{\prime}_2 } & = & {\dst [ \psi _1  - i \, \psi _2 ]/\sqrt{2} } \\
& = & {\dst
(k_x - ik_y)/\sqrt{2}
} & {\dst \mathrm{for}~~\Gamma _5^- }\\
& = & {\dst
k_yk_z (k_x - ik_y)( k_y^2 - 3k_x^2 )/\sqrt{2}
} & {\dst \mathrm{for}~~\Gamma _6^- }.
\eac
\ee
The generators for the invariant polynomials of $\Gamma _{5,6}^-$
of $D_{6h}$ are the same as those for $\Gamma _3^-$ of $O_h$ (\ref{genG3Oh}),
and the free energy may be expressed
\be
\barcl
{\dst F(D_{6h}(\Gamma _{5,6}^-))} & = & {\dst \sum _{m,n,p} 
 F_{(m,n,p)}^+ \, P_{(m ,n ,p)}^+ 
(\eta ^{\prime}_1, \eta ^{\prime}_2)
 + F_{(m,n,p)}^- \, P_{(m ,n ,p)}^- 
(\eta ^{\prime}_1, \eta ^{\prime}_2)
} \\ & = & {\dst
\alpha \, (|\eta ^{\prime}_1 |^2 + |\eta ^{\prime}_2 |^2) +
\beta _1\p \, (|\eta ^{\prime}_1 |^2 + |\eta ^{\prime}_2 |^2)^2 +
4\beta _2\p \, |\eta ^{\prime}_1 \eta ^{\prime}_2 |^2 + \cdots }
\eac
\label{FG56D6h}
\ee
This is identical in form to $F(O_{h}(\Gamma _{3}^-))$, so the
phase transitions take place at the same values of $\alpha$,
$\beta _i$, $\gamma _i$, etc.  Of course, the symmetries of the
phases are different in the two cases.  Also, the dependence of
the parameters $\alpha$, $\beta _i, \ldots$ on physical quantities
such as couplings, masses, the temperature and the pressure are
different, so systems with different normal state 
symmetries do not 
sit at analogous locations in the superconducting phase diagram in
general.

Next we consider the
two dimensional representation $\Gamma _5^-$ of $D_{4h}$.  
The action of $D_{4h}$ on this irrep is isomorphic to
$S_2\times {\mathbf{Z}}_2^2$; that
is, it permutes $\eta _1$ and $\eta _2$ and changes their
signs.  In particular,
${S_2} = \{ E, C_{2e}\}$, and
${\mathbf{Z}_2} = \{ E, C_{2y} \}, \{ E, C_{2z} \}$. 
The generators for the invariant polynomials of $\Gamma _{5}^-$
of $D_{4h}$ are
\be
\barcl
{\dst P_1} & = & 
{\dst |\eta _1|^{2} + |\eta _2|^{2} } \\
{\dst P_2} & = & 
{\dst \left[ \left| \eta _1^{2} + \eta _2^{2} \right| ^2 
- \left( |\eta _1|^{2} + |\eta _2|^{2} \right) ^2 \right] 
~=~ 
-4 \, \left[ \Im \left( \eta _1 \etabar _2\right) \right] ^2 } \\
{\dst P_3} & = & 
{\dst 4 \, |\eta _1\eta _2|^{2} } \\
{\dst P_4} & = & 
{\dst 4 \, \left( |\eta _1|^{2} - |\eta _2|^{2} \right) ~
\Im \left[ (\eta _1 \etabar _2)^{2} \right] } 
\eac
\label{genG5D4h}
\ee
The basis elements are generated exactly as in the case of 
$\Gamma _3^-$ of $O_h$ (\ref{genBasisG3Oh}), and the free
energy is given by
\be
\barcl
{\dst F(D_{4h}(\Gamma _{5}^-))} & = & {\dst \sum _{m,n,p} 
 F_{(m,n,p)}^+ \, P_{(m ,n ,p)}^+ 
(\eta _1, \eta _2)
 + F_{(m,n,p)}^- \, P_{(m ,n ,p)}^- 
(\eta _1, \eta _2)
} \\ & = & {\dst
\alpha \, (|\eta _1 |^2 + |\eta _2 |^2) +
\beta _1 \, (|\eta _1 |^2 + |\eta _2 |^2)^2 -
4 \beta _2 \, \left[ \Im (\eta _1 \etabar _2 ) \right] ^2 + 
4 \beta _3 \, |\eta _1 \eta _2 |^2 + \cdots }
\eac
\ee
The coefficients agree with those used by Sigrist and Ueda \cite{SU}
to fourth order, except for $\beta _3$ which differs by a factor of 4
with theirs larger.

The final irreps are 
the three dimensional representations $\Gamma _4^-$ 
and $\Gamma _5^-$ of $O_h$.  
The action of $O_h$ on each of these irreps is isomorphic to
$S_3\times {\mathbf{Z}}_2^3$; that
is, it permutes $\eta _1, \eta _2$ and $\eta _3$ and changes any of
the signs.  Of course, the correspondence between specific group elements 
and these transformations differs in the two cases.
In $\Gamma _4^-$, $S_3$ is generated by $C_{3\delta}$ and $C_i C_{2f}\p$
and the three copies of ${\mathbf{Z}}_2$ are generated by the
reflections $\sigma _x, \sigma _y$ and $\sigma _z$.
On the other hand, in $\Gamma _5^-$, 
$S_3$ is generated by $C_{3\delta}$ and $ C_{2b}\p$
and the ${\mathbf{Z}}_2$ actions are generated by the
reflections $\sigma _y, \sigma _z$ and $\sigma _x$.
The details of the construction of these invariant 
polynomials and those for the two dimensional irreps
are presented elsewhere \cite{math}.
The generators for the invariant polynomials of $\Gamma _{4,5}^-$
of $O_h$ are
\be
\barcl
{\dst P_1} & = & 
{\dst |\eta _1|^{2} + |\eta _2|^{2} + |\eta _3|^{2}  
\hspace{3.6cm} (SO(6)) } \\ [2mm]
{\dst P_2} & = & 
{\dst |\eta _1 \eta _2|^{2} + |\eta _2 \eta _3|^{2} +
 |\eta _1 \eta _3|^{2} 
\hspace{2.5cm} (U(1)^3\times O_h) } \\ [2mm]
{\dst P_3} & = & 
{\dst |\eta _1 \eta _2 \eta _3|^{2} 
\hspace{5.4cm} (U(1)^3\times O_h) } \\ [2mm]
{\dst P_4} & = & 
{\dst \left| \eta _1^{2} + \eta _2^{2} + \eta _3^{2} \right| ^2
\hspace{4.25cm} (U(1)\times SO(3)) } \\ [2mm]
{\dst P_5} & = & 
{\dst 
\left|  \eta _1^2 \etabar _2^2 + \eta _2^2 \etabar _3^2 + 
\eta _3^2 \etabar _1^2 \right| ^2 - P_2^2
} 
\\ [2mm]
{\dst P_6} & = & 
{\dst \Re \left[ 
|\eta _1|^2 \left( \eta _2^2 \etabar _3^2 - 
|\eta _2 \eta _3|^2 \right)  \right] + cyc.
} \\ [2mm]
{\dst P_7} & = & 
{\dst \Im \left[ 
\eta _1^2 \etabar _2^2 \left( |\eta _1|^2 - |\eta _2|^2 \right) \right] 
+ cyc.
} \\ [2mm]
{\dst P_8} & = & 
{\dst \Re \left[ 
|\eta _1|^2 \eta _2^2 \etabar _3^2 
\left( 2|\eta _1|^2 - |\eta _2|^2 - |\eta _3|^2 \right)  
\right] + cyc. } \\ [2mm]
{\dst P_9} & = & 
{\dst \Im \left[ 
\eta _1^4 \etabar _2^2 \etabar _3^2 + 
\eta _2^4 \etabar _1^2 \etabar _3^2 + 
\eta _3^4 \etabar _1^2 \etabar _2^2  
\right] } \\ [2mm]
{\dst P_{10}} & = & 
{\dst \Im \left[ 
\eta _1^2 \etabar _2^2 |\eta _3 |^2 \left( |\eta _1|^2 - |\eta _2|^2 \right)
\right] + cyc. } \\ [2mm]
{\dst P_{11}} & = & 
{\dst \Re \left[ 
\left( \eta _1^4 \etabar _2^2 \etabar _3^2 
 - |\eta _1^2 \eta _2 \eta_3|^2 \right) 
\left( |\eta _2|^2 + |\eta _3|^2 \right)
\right] + cyc. } \\ [2mm]
{\dst P_{12}} & = & 
{\dst \Im \left[ 
\eta _1^4 \etabar _2^2 \etabar _3^2 \left( |\eta _2|^2 + |\eta _3|^2 \right) 
\right] + cyc. } \\ [2mm]
{\dst P_{13}} & = & 
{\dst \Im \left[ 
\eta _1^2 \etabar _2^2 |\eta _3 |^4 \left( |\eta _1|^2 - |\eta _2|^2 \right)
\right] + cyc. } 
\eac
\label{polyD3}
\ee
where ``cyc.'' denotes additional terms with the indices cyclically 
permuted and we have noted that 
four of the generators have extra continuous symmetries.
Also, $P_5,P_9,P_{11}$ and $P_{12}$ have a ${\mathbf{Z}}_6$ symmetry
in addition to
the requisite $U(1)\times O_h$ for fixed $r=P_1^{1/2}$.
An arbitrary $O_h(\Gamma_{4,5}^-)$ invariant polynomial may be expressed
in terms of the generators (\ref{polyD3}) 
\be
P( \eta _1, \eta _2 , \eta _3 ) =
\sum _{n_1,\ldots ,n_5,X} C_{n_1,\ldots ,n_5}^{(X)} \, 
P_{n_1,\ldots ,n_5}^{(X)} ( \eta _1, \eta _2 , \eta _3 )
\ee
where the basis is given by
\be
P_{n_1,\ldots,n_5}^{(X)} =
P_1^{n_1} \, P_2^{n_2} \, P_3^{n_3} \, P_4^{n_4} \, P_5^{n_5} \, P_X
\ee
where $P_X = 1, P_6, P_6^2, P_6 P_7, P_6^3, P_7, P_8, \ldots, P_{13}$.
This basis has not been constructed previously.

The free energy for $O_h(\Gamma _{4,5}^-)$ is expressed in terms of the
invariant polynomials (\ref{polyD3}):
\be
\barcl
{\dst F} & = & {\dst
\sum _{n_1,\ldots ,n_5,X}
F_{n_1,\ldots ,n_5}^{(X)} \, P_{n_1,\ldots ,n_5}^{(X)}
} \\ & = & {\dst
\alpha P_1 + \beta _1 P_1^2 + \beta _2 P_4 +
\beta _3 P_2 + 
\gamma _1 P_1^3 +
\gamma _2 P_1 P_4+
\gamma _3 P_1 P_2+
\gamma _4 P_3 +
\gamma _5 P_6 +
\gamma _6 P_7 + \cdots 
}
\eac
\label{Fcube}
\ee
where the sum runs over the indices described above. 
These coefficients agree with those used by Sigrist and Ueda 
\cite{SU} to fourth order, and they do not consider the free
energy for $\Gamma _{4,5}^-$ of $O_h$ at higher order.

\subsection{Minimization of the Free Energy}
\label{subsec-Fmin}

The physical gap function minimizes the free energy.  The minimum 
determines both the magnitude and the direction of $\Delta$ in
representation space.  The magnitude depends on the parameters
$\alpha, \beta, \gamma, \ldots$ in a complicated fashion.  Fortunately,
its exact value is unimportant.  It is zero above $T_c$, small just
below the second order critical point and possibly large at low
temperatures.  The direction in $\eta$ space is more interesting, 
since it determines
the symmetry of the superconducting phase.  

Theorem \ref{th-crit} in Appendix \ref{app-thms} guarantees 
that regardless of the 
value of the parameters $\alpha , \beta _1\p, \ldots$ at least one pair
of critical points of the free energy lies on each rotational symmetry
axis of the representation space ($\eta$ space).  The theorem does not
say which of these critical points if any is the absolute minimum,
but sufficiently close to the critical point one of them is \cite{foot-MT}.
This is a consequence of the Morse Theory of critical points
combined with an accounting of the critical points of a fourth
order polynomial ($F$) in terms of point group orbits.  Even
as the magnitude of $\Delta$ grows, an intermediate symmetry phase
is the ground state for most values of the parameters.

\subsubsection{1D Irreps}
 
Because of its simple form, the minimization of the free energy for
one dimensional representations (\ref{FoneDim})
is straightforward in principle.
To fourth order, the free energy is given by
\be
F = \alpha \, |\eta _1|^2 +
\beta \, |\eta _1|^4 
 + \cdots 
\ee
For $\alpha \sim (T-T_c) <0$ and $\beta >0$, the minimum is 
$\eta _1^0 = r_0 \, e^{2i\theta}$ with
\be
r_0 = \sqrt{\frac{|\alpha|}{2\beta }} + \cdots
\ee
where $\theta$ is an arbitrary phase angle which parameterizes the
ground state degeneracy.  This breaks the $U(1)$ symmetry to ${\mathbf{Z}}_2$,
but it does not break the point group symmetry.

Higher terms in the free energy may be considered as well.  
The Ginzburg-Landau formulation is best near the second order 
critical point
where it is well-known that the fourth order free energy 
(perturbed by a few higher order terms to break the residual degeneracy)
provides a good description of the system.  At lower temperatures,
the magnitude of the gap grows and
higher order terms become important.  Eventually the perturbative
expansions in $\Delta /\omega _c$ and $V_{\vec{k},\vec{k}\p}$ may break
down due to a finite radius of convergence and asymptoticity, respectively.
Also, the exact form of the temperature dependence of the coefficients
becomes important, so it no longer suffices to make the ansatz that the 
coefficients are independent of temperature except for $\alpha \sim (T-T_c)$.  
Nevertheless, there does exist an effective free energy even at low
temperatures which is related to the perturbative Ginzburg-Landau 
free energy through resummation, and information about the system at
low temperatures (especially the symmetries) can be extracted from 
the higher order terms.

As these terms are considered, the 
magnitude of the gap function $r_0(\alpha, \beta, \ldots )$ 
takes values on a branched cover of the parameter space. Consider the
minimization of the sixth order free energy, 
$F = \alpha \, |\eta _1|^2 + \beta \, |\eta _1|^4 + \gamma \, |\eta _1|^6 
 + \cdots $,
which is of interest in the case of higher dimensional irreps.  
The minimum becomes
\be
r_0 = \sqrt{\left( -\sgn (\beta ) + 
\sqrt{1+3|\alpha |\gamma /\beta ^2} \right) 
\frac{|\beta|}{3\gamma}} + \cdots
\label{r0six}
\ee
The value of the free energy at the minimum is conveniently expressed
in terms of the function
\be
\barcl
{\dst \Phi (x,y) } &\equiv & {\dst
\frac{-1}{27 |x| y^2 } \left\{ 
2 ( 1 + 3y )^{3/2} - \sgn ( x ) ( 2 + 9y) 
\right\} 
} \\ & = & {\dst
\frac{-1}{4x} \left( 1 - \txthalf y + {\tst \frac{9}{16}} y^2 
- \cdots \right) ~~~~~~~ x>0, |y|<{\tst \frac{1}{3}}
}
\eac
\label{phiDef}
\ee
where the series is its critical (small $\alpha$) expansion.
The free energy is given by
\be
F_{min} = \alpha ^2 \Phi (\beta , |\alpha | \gamma / \beta ^2 ).
\ee
The function $\Phi (x,y)$ is convenient because it is a monotonically
increasing function of both $|x|$ and $y$.  

It is possible to have first order phase transitions in this parameter
space at eighth order, where a small change in one parameter causes a 
large change in the gap function because the minimum
hops from one sheet to another.  The location of these critical
points is non-universal. 
The symmetry of the superconducting
state need not change at this type of phase transition, and we do not
consider them further.  
Note that away from the transitions, the higher order terms
simply renormalize the leading non-zero coupling $\beta$ according
to (\ref{phiDef}), and we define 
$\beta _R \equiv -1/(4\Phi(\beta , |\alpha | \gamma / \beta ^2 ))$
so that $F_{min} = -\alpha ^2/(4\beta _R)$.

\subsubsection{2D Irreps}

The free energy for the two dimensional irreps has the same general
features, but it also allows the point group symmetry to be broken.
We express  $F$
in terms of one real and one complex variable
\be
r^2 = |\eta ^{\prime}_1|^{2} + |\eta ^{\prime}_2|^{2} 
\hspace{1cm} e^w = \frac{\eta _1^{\prime}}{\eta _2 ^{\prime}}
\ee
and conversely,
\be
\eta _1\p = r e^{2i\theta +w/2} \sech ^{1/2} \left( \Re \, w \right)/\sqrt{2}
\hspace{1cm}
\eta _2\p = r e^{2i\theta -w/2} \sech ^{1/2} \left( \Re \, w \right)/\sqrt{2}.
\ee
The remaining degree of freedom in $\eta ^{\prime}_1$ and 
$\eta ^{\prime}_2$ is a phase, $\theta$.  This is the zero mode of
the Goldstone boson which is irrelevant by gauge invariance.
These variables behave nicely under $O_h\times U(1)$.  
For example, in  $\Gamma _3^-$ of $O_h$,  
$C_{3\alpha}: w \ra w + 2\pi i /3$ and
$C_{4z}: w \ra -w$.
Consider the two dimensional irreps $\Gamma _3^-$ of $O_h$ and 
$\Gamma _{5,6}$ of $D_{6h}$ first.  
The generators of the invariant polynomials for these irreps 
(\ref{genG3Oh}) become
\be
\asize{1.7}
\barcl
{\dst P_1} & = & {\dst r^2 } \\
{\dst P_2} & = & 
{\dst r^4 \, \sech ^2 ( \Re \, w) } \\
{\dst P_3} & = & 
{\dst r^6 \, \cos ( \Im \, 3w ) \, 
\sech ^3 ( \Re \, w ) } \\
{\dst P_4} & = & 
{\dst r^8 \, \sin ( \Im \, 3w )
\, \sech ^3 ( \Re \, w ) \, \tanh  ( \Re \, w ) } 
\eac
\label{gens}
\ee
which are manifestly invariant under the symmetry operations.
In the $(r,w)$ variables, the free energy is 
\be
\asize{1.4}
\barcl
{\dst F} & = & {\dst \alpha \, r^2 + \beta _1\p \, r^4 +
\beta _2\p \, r^4 \sech ^2 ( \Re \, w) 
} \\ & & \hspace{1cm} {\dst
+ \gamma _1 \, r^6 + \gamma _2 \, r^6 \, \sech ^2 (\Re \, w)
+ \gamma _3 \, r^6 \, \cos ( \Im \, 3w ) \, \sech ^3 ( \Re \, w )
+ \cdots } 
\eac
\ee
with the same coefficients defined in Eq. (\ref{FG3Oh}).

The order parameter $\Delta (r,w)$ 
is non-zero below the critical point, breaking the
$U(1)$ symmetry.  The value of $w$ (the direction of $\Delta$ in
$\eta$ space) below $T_c$ determines whether and 
how the point group is broken. 
A glance at the functions $P_1, \ldots P_4$ (\ref{gens})
which generate the invariant polynomials reveals that the generic
extrema are $w= +\infty , -\infty ,0,i\pi /3,\dots ,5\pi i/3$.  
These are the critical points identified by Theorem 2 in 
Appendix \ref{app-thms},
since the rotational symmetry axes in the two dimensional $\eta \p$ 
representation space are $(1,0),(0,1),(1,\pm 1),(1,\pm \omega)$ and
$(1,\pm \omega^2)$.
They give a non-trivial residual symmetry.

The minimization of the free energy is straightforward. 
We consider $F$ to sixth order. 
The imaginary part
of $w$ only enters through $\cos ( \Im \, 3w )$, so 
\be
\Im \, w = 
\left\{ 
\barcl
0 ,{\tst \frac{2\pi}{3}},{\tst \frac{4\pi}{3}} 
~~~~ \gamma _{3R} & < & 0 \\
{\tst \frac{\pi}{3}}, \pi, {\tst \frac{5\pi}{3}} ~~~~~ \gamma _{3R} & > & 0
\eac
\right.
\label{imw2}
\ee
where the renormalized coupling $\gamma _{3R}=\gamma _3$ at sixth order.
At this order
of perturbation theory, $\Im \, w$ is unconstrained if $\gamma _3 =0$.
The free energy takes the form $F = 
\alpha \, r^2 + \beta _1\p \, r^4 + \beta _2\p \, r^4 x^2 
+ \gamma _1 \, r^6 + \gamma _2 \, r^6 \, x^2 
- |\gamma _3| \, r^6 \, x^3$ to sixth order with  
$\Im \, w$ given by (\ref{imw2}) and $x=\sech ( \Re \, w) \in [0,1]$.  
The minima either have $x=0$ or $x=1$, since the coefficient of $x^3$
is negative and there is no term proportional to $x$. 
They correspond to phases in which the point group is only partially
broken, and $\Delta$ points along a rotational symmetry axis in
$\eta$ space.  Specifically, these phases are
\be
\barll
{\dst
x = 0:
} \\ & \bal
{\dst 
\Delta ~ = ~ r_0 k_xk_yk_z ( k_z^2 + \omega k_x^2 + \omega ^2 k_y^2 )
} \\ {\dst
\Delta ~ = ~ r_0 k_xk_yk_z ( k_z^2 + \omega ^2 k_x^2 + \omega k_y^2 )
} \eac
 & {\dst \mathrm{for}~~\Gamma _3^-~of~O_{h} }\\
 & \bal {\dst
\Delta ~ = ~ r_0 ( k_x \pm i k_y ) 
} \eac & {\dst \mathrm{for}~~\Gamma _5^-~of~D_{6h} }\\
 & \bal {\dst
\Delta ~ = ~  r_0 k_yk_z (k_x \pm ik_y)( k_y^2 - 3k_x^2 )
} \eac ~~~~~~~~~~~  
 & {\dst \mathrm{for}~~\Gamma _6^-~of~D_{6h} } \\ [2mm]
{\dst
x = 1:
} \\  & \balr
{\dst \Delta ~ = ~ r_0 k_xk_yk_z ( 2 k_z^2 -  k_x^2 - k_y^2 ), cyc. ~~~} 
 & \hspace{0.2cm} {\dst (\gamma _3 < 0) } \\ 
{\dst \Delta ~ = ~ \sqrt{3} r_0 k_xk_yk_z ( k_x^2 - k_y^2 ), cyc. }
 & \hspace{0.2cm} {\dst (\gamma _3 > 0) } \eac  ~~~~~~~
 & {\dst \mathrm{for}~~\Gamma _3^-~of~O_{h} }\\
  & \balr
{\dst \Delta ~ = ~ r_0 k_x, 
r_0 ( \txthalf k_x \pm {\tst \frac{\sqrt{3}}{2}} k_y ) ~~~} 
 & \hspace{1.7cm} {\dst (\gamma _3 < 0) } \\ 
{\dst \Delta ~ = ~ r_0 k_y, 
r_0 ( \txthalf k_y \pm {\tst \frac{\sqrt{3}}{2}} k_x )  ~~~} 
 & \hspace{1.7cm} {\dst (\gamma _3 > 0) } \eac
 & {\dst \mathrm{for}~~\Gamma _5^-~of~D_{6h} }\\
 & \balr 
{\dst \Delta ~ = ~ r_0 k_xk_yk_z ( k_y^2 - 3k_x^2 ),\cdots ~~~} 
  & \hspace{1.3cm} {\dst (\gamma _3 < 0) } \\ 
{\dst \Delta ~ = ~ r_0 k_y^2k_z ( k_y^2 - 3k_x^2 ), \cdots ~~~} 
  & \hspace{1.3cm} {\dst (\gamma _3 > 0) } \eac
& {\dst \mathrm{for}~~\Gamma _6^-~of~D_{6h} } 
\eac
\ee
where $r_0$ is the value of
$r$ for the ground state, and ``{\em cyc.}'' means terms of the same
form with the indices cyclically permuted.  

The magnitude of the ground state gap function, $r_0$, is given by
an expression of the same form as Eq.\ (\ref{r0six}), with 
$\beta = \beta _1\p$ and $\gamma = \gamma _1$ for $x=0$ and
$\beta = \beta _1\p + \beta _2\p$ and
$\gamma = \gamma _1 + \gamma _2 - |\gamma _3|$, for $x=1$. 
It is natural to define the following renormalized couplings
\be
\barcl
{\dst \beta _{1R} 
} & = & {\dst
\frac{-1}{4 \Phi ( \beta _1\p , |\alpha | \gamma _1 / \beta _1^{\prime \, 2})}
} \\
{\dst \beta _{2R} 
} & = & {\dst
-\beta _{1R} + 
\frac{-1}{4 \Phi ( \beta _1\p +\beta _2\p, 
|\alpha | (\gamma _1 + \gamma _2 - |\gamma _{3R}|)
/(\beta _1\p +\beta _2\p )^2)}
}
\eac
\ee
so that 
the free energy of the ground state is given by
$F_{min}(x=0) = -\alpha ^2 /(4\beta _1\p)$ and
$F_{min}(x=1) = -\alpha ^2 /(4(\beta _1\p+\beta _2\p))$.
The ground state is 
$x=0$ when $\beta _{2R} >0$ and
$x=1$ when $\beta _{2R} <0$.
The ground state gap functions for the three distinct phases are
shown in Table \ref{table-phases}.
Typically, the
value of $r_0$ changes discontinuously at the boundary between
these phases, so the transition is first order, as expected when
the symmetry does not break to a subgroup.

At eighth and higher orders, there are small regions of the parameter space
in which the ground state has $0<x<1$ and the point group is broken 
completely: 
$\Gamma _3^-(O_h) \ra \Gamma _1^-(D_{2h}),
\Gamma _5^-(D_{6h}) \ra \Gamma _1^-(C_{i})$ and 
$\Gamma _6^-(D_{6h}) \ra \Gamma _1^-(C_{i})$, where $C_i$ is inversion.
Second order transitions to these phases can result from frustration
due to the competition between the terms minimized at $x=0$ and those
minimized at $x=1$ (or from the contribution of the symmetry breaking
generator $P_4$).  Consider the eighth order free energy.  It has the
terms $\beta _2\p r^4 x^2$ and $\delta _4 r^8 x^4$ which compete
when $\beta _2\p < 0$ and $\delta _4 >0$.  The free energy is a
fourth degree polynomial in $x$, which decreases as $x$ increases from
$0$, then reaching a minimum, it increases.  If the minimum occurs
at $x>1$, then the ground state is $x=1$ since $x=\sech \Re \, w \le 1$.
On the other hand, if the minimum occurs at $x<1$, the ground state
is the low symmetry phase.  We can estimate the critical point
taking $|\beta _{2R}|\ll \beta _{1R}$, for which the low symmetry
phase exists when 
\be
|\alpha | \ge \left( \frac{4|\beta _{2R} | \beta _{1R}^2}{\delta _4}
\right) ^{1/2}
\ee
with $\beta _{2R} <0$ and $\delta _4>0$.  
Since $\alpha \sim (T-T_c)/T_c$ is small near the initial
superconducting critical point, this secondary phase transition
would occur at much lower temperatures.

The other two dimensional irrep is $\Gamma _5^-$ of $D_{4h}$.
Again, the simple groups act nicely on the projective variables,
$C_{2y}: w \ra w + i\pi $ and
$C_{2e}: w \ra -w$.
The rotational symmetry axes in the two dimensional representation
space are $(1,0),(0,1),(1,\pm 1)$ and $(1,\pm i)$, which 
correspond to $w=\infty, -\infty, 0,i\pi ,i\pi /2$ and $-i\pi /2$.
The generators of the invariant polynomials are
\be
\asize{1.4}
\barcl
{\dst P_1} & = & {\dst r^2 } \\
{\dst P_2} & = &
{\dst -r^4 \, \sin ^2 ( \Im \, w ) \,
\sech ^2 ( \Re \, w )} \\
{\dst P_3} & = &
{\dst r^4 \, \sech ^2 ( \Re \, w) } \\
{\dst P_4} & = &
{\dst r^6 \, \sin ( \Im \, 2w ) 
\, \sech ^2 ( \Re \, w ) \,
\tanh  ( \Re \, w ) }
\eac
\label{G5D4hgens}
\ee
which are manifestly invariant.
The free energy becomes
\be
\barcl
{\dst F} & = & {\dst \alpha \, r^2 + \beta _1 \, r^4 
 - \beta _2  \, r^4 \, \sin ^2 ( \Im \, w ) \,
\sech ^2 ( \Re \, w )
 + \beta _3 \, r^4 \sech ^2 ( \Re \, w) 
} \\ & & \hspace{1cm} {\dst
+ \gamma _1 \, r^6 
- \gamma _2 \, r^6 \, \sin ^2 ( \Im \, w ) \, \sech ^2 ( \Re \, w )
+ \gamma _3 \, r^6 \, \sech ^2 (\Re \, w)
} \\ & & \hspace{1cm} {\dst
+ \gamma _4 \, r^6 \, \tanh  ( \Re \, w )
\, \sech ^2 ( \Re \, w ) \, \sin ( \Im \, 2w ) + \cdots }
\eac
\ee
The structure of the
solution is more complicated than in the case of the other two
dimensional representations because the generator $P_4$ that explicitly
breaks the intermediate symmetries occurs at sixth order rather than
eighth order.  This leads to low temperature phases in which the point 
group is broken to $\II$.  

We restrict our attention to the sixth
order free energy with no symmetry breaking parameter: $\gamma _4=0$.
The relative phase for the ground state is given by
\be
\Im \, w = 
\left\{ 
\ball
0 , \pi
& \beta _{2R} ~ < ~ 0 \\
{\tst \frac{\pi}{2}}, {\tst \frac{3\pi}{2}} ~~~~~ & \beta_{2R} ~ > ~ 0
\eac
\right.
\label{imw2a}
\ee
with $\beta_{2R} = \beta _2  + \cdots$ (See Eq. (\ref{betaTwoR})\ ).
The free energy reduces to $F(\Im \, w=\pi/2) =
\alpha \, r^2 + \beta _1 \, r^4 - \beta _2 \, r^4 x^2 
+ \beta _3 \, r^4 x^2
+ \gamma _1 \, r^6 - \gamma _2 \, r^6 \, x^2 
- \gamma _3 \, r^6 \, x^2$, and $F(\Im \, w=0)$ given by
the same expression with $\beta _2 = \gamma _2=0$. Note that
the free energy is quadratic in $x$.  The phase boundaries are
again best expressed in terms of renormalized couplings,
\be
\barcl
{\dst \beta _{1R} 
} & \equiv & {\dst
\frac{-1}{4 \Phi ( \beta _1 , |\alpha | \gamma _1 / \beta _1^{2})}
} \\
{\dst \beta _{3R} 
} & \equiv & {\dst
-\beta _{1R} + 
\frac{-1}{4 \Phi ( \beta _1 +\beta _3, 
|\alpha | (\gamma _1 + \gamma _3 )
/(\beta _1 +\beta _3 )^2)}
} \\
{\dst \beta _{2R} 
} & \equiv & {\dst
\beta _{1R} + \beta _{3R} + 
\frac{1}{4 \Phi ( \beta _1 -\beta _2 +\beta _3, 
|\alpha | (\gamma _1 - \gamma _2 + \gamma _3 )
/(\beta _1 -\beta _2 +\beta _3 )^2)}
}
\eac
\label{betaTwoR}
\ee
to sixth order, so that 
$F_{min} = -\alpha ^2 /(4\beta )$ with 
$\beta = \beta _{1R}$ for $w=\pm \infty$, 
$\beta = \beta _{1R}+ \beta _{3R}$ for $w=0,i\pi $, 
$\beta = \beta _{1R}-\beta _{2R}+ \beta _{3R}$ for $w=\pm i\pi /2$. 
The ground states are 
$w=0,i\pi $ when $\beta _{2R},\beta _{3R} <0$,
$w=\pm \infty$ when $\beta _{3R}> \max (0,\beta _{2R})$,
and
$w=\pm i\pi /2$ when $\beta _{2R}> \max (0,\beta _{3R})$.
Expressions for the gap function
in terms of the original variables for each of the three phases are
given in Table \ref{table-phases}.

\subsubsection{3D Irreps}

In analogy with the analysis of the two dimension irreps, the free
energy for the three dimensional irreps $\Gamma _4^-$ and
$\Gamma _5^-$ of $O_h$ should be expressed in terms of complex
projective coordinates.  There are two relative magnitudes and
two relative phases, so $\eta _1, \eta _2$ and $\eta _3$ should
be expressed in terms of a two dimensional vector in complex 
projective space
$\vec{w}\in {\mathbf{CP}}^2$; however, there is no known vector
that behaves simply under the $O_h$ operations.  The best we can
do is to use the homogeneous coordinates, $\eta _j/(r e^{2i\theta})$,
and there is no apparent simplification of the free energy.

The ground state of the free energy to sixth order is given by
a generic critical point provided the symmetry breaking parameter
$\gamma _6$ is taken to be zero.  The generic critical points are
$(0,0,1),(1,1,0), (1,1,1), (1,i,0), (1,\omega ,\omega ^2)$ and
$(1,\omega ,\omega ^2)$ up to symmetry, according to 
Theorem 2 of Appendix \ref{app-thms}.
At these points,
the invariant polynomial generators (\ref{polyD3}) have the following values
\begin{center}
\asize{1.2}
\begin{tabular}{lccccccc} 
$~~~\hat{\eta}$ & ~~$P_1$~~ & ~~$P_2$~~ & ~~$P_3$~~ & ~~$P_4$~~ 
& ~~$P_5$~~ & ~~$P_6$~~ & ~~$P_{7-13}$~~ \\
$(0,0,1)$ & 1 & 0 & 0 & 1 & 0 & 0 & 0 \\
$(1,1,0)$ & 1 & $\frac{1}{4}$ & 0 & 1 & 0 & 0 & 0 \\
$(1,1,1)$ & 1 & $\frac{1}{3}$ & $\frac{1}{27}$ & 1 & 0 & 0 & 0 \\
$(1,\omega ,\omega ^2)$ & 1 & $\frac{1}{3}$ & $\frac{1}{27}$ & 0 & 
   0 & $-\frac{1}{6}$ & 0 \\
$(1,i,0)$ & 1 & $\frac{1}{4}$ & 0 & 0 & 0 & 0 & 0 
\end{tabular}
\asize{1.0}
\label{genVals}
\end{center}
at $r=1$.
Let $\beta _{1R}\equiv 
-1[4 \Phi ( \beta _1 , |\alpha | \gamma _1 / \beta _1^{2})]^{-1}$, and
define the other renormalized couplings such that 
the free energy of each phase is given by
\be
\barcl
{\dst F_{(0,0,1)} 
} & = & {\dst
\frac{-\alpha ^2}{4(\beta _{1R} + \beta _{2R})}
} \\
{\dst F_{(1,i,0)} 
} & = & {\dst
\frac{-\alpha ^2}{4(\beta _{1R} + \frac{1}{4} \beta _{3R})}
} \\
{\dst F_{(1,1,1)} 
} & = & {\dst
\frac{-\alpha ^2}{4(\beta _{1R} + \beta _{2R}+ \frac{1}{3} \beta _{3R})}
+ \frac{\gamma _{4R}}{27} 
\left( \frac{|\alpha |}
 {2(\beta _{1R} + \beta _{2R}+ \frac{1}{3} \beta _{3R})} \right) ^3} \\
{\dst F_{(1,\omega ,\omega ^2)} 
} & = & {\dst
\frac{-\alpha ^2}{4(\beta _{1R} + \frac{1}{3}\beta _{3R})}
+ \left( \frac{\gamma _{4R}}{27} - \frac{\gamma _{5R}}{6} \right)
\left( \frac{|\alpha |}
 {2(\beta _{1R} + \frac{1}{3} \beta _{3R})} \right) ^3} \\
{\dst F_{(1,1,0)} 
} & = & {\dst
\frac{-\alpha ^2}{4(\beta _{1R} + \beta _{2R} + \frac{1}{4} \beta _{3R})}
+ \cdots
} 
\eac
\ee
where the couplings are defined sequentially, $\beta _{2R}$ by $F_{(0,0,1)}$, 
then $\beta _{3R}$ by $F_{(1,i,0)}$,
then $\gamma _{4R}$ by $F_{(1,1,1)}$ and finally 
$\gamma _{5R}$ by $F_{(1,\omega ,\omega ^2)}$. 
The free energies may be expressed in terms of the original (bare) couplings
through the $\Phi$ function as before, using the coefficients listed above. 
Note that $\beta _{jR} = \beta _j$ to fourth order,
but the renormalized $\gamma$'s receive corrections even at
sixth order, $\gamma _{4R}=\gamma _4 + \cdots$ and 
$\gamma _{5R}=\gamma _5 + \cdots$.

The ground state of the cubic system is shown in Table \ref{table-phases}, 
and the phase diagram is shown in Fig.~\ref{fig-cubPhase} taking
$\gamma _{4R}=\gamma _{5R} =\gamma _{6R} =0$.  
When $\gamma _{4R}$ and $\gamma _{5R}$ are
non-zero, the phase boundaries shift by a small amount.
They also have a more interesting qualitative effect in that they
allow the possibility of stabilizing $(1,1,0)$ as the 
ground state.  At fourth order, 
even though $(1,1,0)$ is a rotational symmetry axis,  
the $(1,1,0)$ phase only exists on
the negative $\beta _{2R}$ axis (see Fig. \ref{fig-cubPhase}) where
it is degenerate with the $(0,0,1)$ and the $(1,1,1)$ phases.  This
degeneracy is the result of the enhanced $SO(3)\times U(1)$ symmetry 
(see Appendix \ref{app-O3}).
The fact that the $\hat{\eta} = (1,1,0)/\sqrt{2}$ 
is only a saddle point at fourth order and not a minimum follows from 
Morse Theory 
constraints.  This is no longer true in the larger parameter
space of the sixth order free energy, and the ground state is allowed to 
be $(1,1,0)$. Taking $\beta _3=0$, we find this occurs for 
$\gamma _4 > -\frac{27}{12}\gamma _3 >0$.  

\subsection{Weak Coupling}

This completes the construction of the generalized phase diagrams for
SSS with cubic, hexagonal and tetragonal symmetry in the normal phase.  
We are now in a position to reexamine the superconducting phase 
transition in the weakly coupled theory.  
Using standard techniques \cite{VW}, we find that the weak coupling
expansion of the free energy for $\Delta$ in an irrep of $G$ 
is given by 
\be
\barcl
{\dst F } & = & {\dst F_{N} + 
\frac{1}{2} N(0) \log \frac{T}{T_c} 
\langle | \Delta (\vec{v}_k) |^2 \rangle _{\khat} +
\sum _{m=1}^{\infty} 
F_{2m+2} \, \langle | \Delta (\vec{v}_k) |^{2m+2} \rangle _{\khat} }~, \\
 & & {\dst  ~~~~~F_{2m+2} =
-\frac{1}{2} \frac{(2m-1)!!}{(m+1)!} \left( 2^{2m+1} -1 \right)
 \zeta (2m+1) \left( \frac{-\beta _c^2}{8 \pi ^2} \right) ^m N(0) }
\eac
\ee
where $F_N$ is the free energy of the normal phase and 
$\beta _c = 1/k_BT_c$.  Note that weak coupling means that the
parameter $\beta _c^2 |\Delta |^2_{RMS}$  is small.
As in Sec.~\ref{sec-Pairing}.B, the simplest ansatz is to consider
an interaction of the form (\ref{kxkp})
\be
W_{\vec{k},\vec{k}\p} =
  -|{\WW}|\frac{\vec{v}_k \cdot \vec{v}_{k\p}}{v_F^2/3},
\label{FSHint}
\ee
where we have rewritten it in terms of the Fermi velocity appropriate
for non-spherical Fermi surfaces.  The coefficients
in the free energy (\ref{Fcube}) are then integrals over 
Fermi surface harmonics \cite{pba}, which are found to equal
$\alpha = \frac{1}{6} N(0) \log (T/T_c), 
\beta _1 = \frac{2}{15} F_4, 
\beta _2 = \frac{1}{15} F_4, 
\gamma _1 = \frac{2}{35} F_6, 
\gamma _2 = \frac{3}{35} F_6$ and 
$\beta _3= \gamma _3 = \gamma _4 = \gamma _5 = 0$ 
assuming a spherical Fermi surface.  In this case,
the free energy reduces to the
form invariant under $SO(3)\times U(1)$ which is studied in Appendix
\ref{app-O3}, and the weakly coupled system sits at the point 
on the positive $\beta _{2R}$ axis near the origin, denoted 
by an ``$\times $'' on the phase
diagram in Fig.~\ref{fig-cubPhase}.  As the Fermi surface is deformed
outward at the diagonals inducing a positive hexadecapole moment,
$\beta _3$ becomes positive, and $(1,i,0)$ becomes the ground state.
On the other hand, if the Fermi surface is deformed inward,
$\beta _3$ is negative, and $(1,\omega ,\omega ^2)$ is the ground state.
Therefore, we identify $(1,i,0)$ and $(1,\omega ,\omega ^2)$ in
$\Gamma _4^-$ of $O_h$ as the leading candidates for the SSS
ground state in cubic systems.  In hexagonal and tetragonal systems,
the candidates are $\Delta \propto (1,i,0) \cdot \vec{v}_k$ in $\Gamma _5^-$
of $D_{6h}$ and $D_{4h}$.

\section{Thermodynamic Properties}
\label{sec-Thermo}
\setcounter{equation}{0}

The gap function 
$\Delta _{\vec k} $ may be zero for certain values of $\vec k$.  
This can occur accidentally, but in some cases the gap function is 
required to vanish by symmetry. Such zeros are robust and
have a marked impact on the
low temperature behavior of thermodynamic quantities such as the heat
capacity and acoustic attenuation.  The characteristic 
$e^{-2\Delta /kT}$ exponential 
behavior due to the finite gap, changes to a power law
behavior when the gap function has zeros.  

The gap function for SSS always has zeros.  This is a consequence of its
odd parity and the fact that it has a single spin component.  The zeros 
are guaranteed by topological considerations very reminiscent of the
so-called ``Hairy Ball'' Theorem (the Hopf or Poincar\'e-Hopf Theorem)
which relates the total index of a vector field to the Euler
characteristic of the underlying (closed, orientable) surface
\cite{DFN}.  Consider the field 
$e^{i\phi} = \Delta / |\Delta |$ where $\phi$ is real, 
which is a well-defined field on the 
Fermi surface assuming that $\Delta$ has no zeros.  
It takes values on the unit circle in the complex plane.
Since $\Delta$ changes sign under inversion,
$\phi (-\vec{k}) = \phi (\vec{k}) + (2k+1) \pi$.  Following the value
of $\phi (\vec{k})$ as $\vec{k}$ is taken around any equatorial circle
on the (closed) Fermi surface, we find that $\phi$ comes back to itself
up to an odd winding number, $\delta \phi = 2 \pi (2k+1)$.  The non-zero
winding number prevents the equatorial circle from being contracted
smoothly to a point on the Fermi surface.  It must encounter a zero
of $\Delta$.  Although the gap function may have zeros in singlet and
triplet superconductivity, only in SSS is it guaranteed to have them.
The gapped states of triplet superconductivity (the Balian-Werthamer
ground state \cite{BW}) do not occur because $\Delta$ only has 
one spin component in SSS.

These generic zeros of the gap function may be found 
using Theorem \ref{th-zero} in Appendix \ref{app-thms}.  
They are given by fixed points of
elements of the residual symmetry group $H$ with a non-trivial 
character, and are tabulated in Table \ref{table-zero}.
Isolated point nodes arise as fixed points of rotations, whereas 
lines of zeros are associated with reflections.  
The nodal structure of $\Delta$ determines the density of
states near the Fermi surface and consequently the scaling 
of thermodynamic quantities.  
With a few assumptions the scaling exponents may be
computed.  For example, the heat capacity, which for a constant gap
vanishes exponentially, 
scales as $T^3, T^2$ and $T$ with point nodes,
line nodes and vanishing order parameter, respectively,
in a defect-free superconductor \cite{VG}.  Multiple line
nodes lead to $T\log T$ \cite{logTerms}.  
This power law scaling is a hallmark
of unconventional superconductivity \cite{VG}.  It must
occur in SSS.  For example, the candidate ground states $(1,i,0)$
and $(1,\omega ,\omega ^2)$
of $\Gamma _4^-$ of $O_h$ discussed above have  point nodes,
so their intrinsic heat capacities scale as $T^3$.




\section{Summary}
\setcounter{equation}{0}

The occurrence of a HM AFM normal state, with one conducting and one
insulating spin channel, has been shown to provide the possibility
of a novel superconducting phase for which the operation of time
reversal has no part.  The state is best considered as a condensed
phase of spinless fermions, with a gap function that is odd in 
$\vec{k}$ and a pair wave function that is odd upon interchange
of particles.  The form of anisotropic interaction required to 
form this state
was obtained, and the resulting gap equation is found to be of
the BCS form.
The allowed symmetries of the gap function have been
enumerated for cubic, hexagonal and tetragonal lattices, and the
corresponding conditions on the parameters of the free energy have been
determined.  The quasiparticle spectrum is necessarily gapless.
For point nodes or a line of nodes, this gaplessness gives 
rise to power-law behavior in T or $\omega$.  For intersecting lines
of nodes, there will be logarithmic terms such as $T\log T$ 
as enumerated by Nazarenko
\cite{logTerms}.

It is anticipated that examples of such phases can be
found in transition metal compounds.  The recent suggestions that 
Sr$_2$RuO$_4$ may be displaying triplet superconductivity\cite{sr2ruo4},
together with predictions of HM AFM states in transition metal 
oxides\cite{WEP2}, both indicate that transition metal oxide compounds
present a favorable possibility of obtaining single spin superconductors.

\vspace{3mm}

\begin{center}
ACKNOWLEDGMENTS
\end{center}

This work was supported by the Office of Naval Research.  

\appendix
\section{Two Theorems}
\label{app-thms}
\renewcommand{\theequation}{\Alph{section}.\arabic{equation}}

The minimum of the free energy determines the order parameter of the
superconducting phase, and the zeros of the gap function in turn
determine the scaling of thermodynamic properties with temperature.
The generic values of these minima and zeros are fixed by the
symmetries of the system, and they may be determined without resorting
to explicit representatives of the symmetry.  

This Appendix presents two theorems which are useful in this regard.
Theorem \ref{th-zero} may be used to find the generic zeros of the 
gap function.  The zeros arise as fixed points of elements of the 
residual symmetry group $H$ that have non-trivial 
character.  This is a refinement of the procedure used by
Volovik and Gor'kov \cite{VG1}, who identified
the zeros with specific group elements.  The utility of our
theorem is that the appropriate group elements may be read
off of standard character tables.

Theorem \ref{th-crit} may be used to find the direction
of the generic critical points of the free energy in representation
space.  Note that the magnitude of the solution (the magnitude of the
gap function) is not determined, but this does not affect the symmetry
of the superconducting phase.  The theorem also applies to more
traditional applications, such as to show that the Fermi surface is
orthogonal to the axes of rotation of the crystal where they intersect.
This theorem applies to the full non-perturbative free energy.

\begin{theorem}
\label{th-zero} 
Suppose $\Delta (\vec{k})$ is in the irrep 
$\Gamma \p$ of the little group $H$, and suppose $g\in H$ has a non-trivial
character, $\chi _{\Gamma \p} (g) \ne 1$.  Then $\Delta (\vec{k}_0)=0$
for any fixed point $\vec{k}_0$ of g (i.e.\ for any $\vec{k}_0$ such that
$\ghat \, \vec{k}_0 = \vec{k}_0$).
\end{theorem}

\noindent
Proof:  Since $H$ is the little group, $\dim \Gamma \p =1$ and
\be
g: \Delta (\vec{k}_0)  \ra \Delta (\ghat \vec{k}_0) ~ = ~
\chi _{\Gamma ^{\prime}} (g) \, \Delta (\vec{k}_0)
\ee
But $\vec{k}_0$ is a fixed point, so
\be
\left( 1 - \chi _{\Gamma ^{\prime}} (g) \right) \Delta (\vec{k}_0) = 0.
\ee
And we arrive at the result, $\Delta (\vec{k}_0) = 0$ provided
$\chi _{\Gamma \p} (g) \ne 1$. $\Box$

This theorem provides a relatively easy means to identify the nodes
of the gap function guaranteed by symmetry.  
As an example, consider the $(1,\omega,\omega ^2)$
state of $\Gamma _4^-$ of $O_h$, which transforms as $\Gamma _2^-$ of 
the little group $C_{3i}$
(see Table \ref{table-phases}).  The characters of the elements of
$C_{3i}$ are listed in common character tables \cite{foot-CG}.
$C_3,C_3^{-1},I,S_6$ and $S_6^{-1}$ have non-trivial characters
and of those, $C_3$ and $C_3^{-1}$ have fixed points, the two points
where the axis of rotation (1,1,1) intersects the Fermi surface.

This information is tabulated in Table \ref{table-zero}.
Note that in some cases any function in the specified irrep must
have a divisor whose little group is larger than the little group
of the function itself.  Theorem \ref{th-zero} applies to these
factors as well, and any zero of the factor is also a zero of the
function.  An example is $k_y^2 k_z (k_y^2-3k_x^2)$ which 
transforms as $\Gamma _6^-$ of $D_{6h}$.  Its little group is
$D_{2h}$, but it has the factor $k_y k_z (k_y^2-3k_x^2)$ that
transforms as the one dimensional irrep $\Gamma _4^+$ of $D_{6h}$.
This accounts for two additional lines of nodes.

\begin{theorem}
\label{th-crit} 
Suppose that $\eta$ transforms 
as the irrep $\Gamma$ of $G$,  a subgroup of $O_h$ or $D_{6h}$, 
where $\eta$ is a $\dim \Gamma$ dimensional complex vector and 
suppose that the $U(1)$ invariant function $F(\eta )$ is in the 
trivial irrep $\Gamma _1^+$ of $G$.  
Also suppose that there exists an element $g\in G$ 
with a fixed point $\eta _0$ 
up to a phase: $\ghat \eta _0 = e^{i\phi} \eta _0$.
Then $e^{i\phi}g$ lies in the little group of the gradient 
$\frac{\d F ( \eta _0 )}{\d \eta _j}$.  
\end{theorem}

Note that $\dim \Gamma \p \le 3$. 
Also note that $\frac{\d F}{\d \eta _j} \equiv \d _j F$ 
is the complex gradient, which
could be expressed in terms of its $2\dim \Gamma$ real components.  The 
theorem states that all of these components are invariant under 
$e^{i\phi}g$ where the phase is equal to the one that appears in the
$\eta _0$ transformation.  

\vspace{2mm} 

\noindent
Proof:  
\be
g: \d _j F ( \eta _0 )  \ra 
\left. \d _{j} F ( \eta  ) \right| _{\ghat \eta _0} ~ = ~
\ghat _{ij} \d _{j} F ( \eta _0 )
\ee
Since $\ghat \eta _0 = e^{i\phi} \,\eta _0$, we have
\be
e^{i\phi}\, \ghat _{jk} \d _{k} F ( \eta _0 ) = \d _{j} F ( \eta _0 )
\ee
Thus, $\d _{j} F ( \eta _0 )$ is invariant under $e^{i\phi}g$;
that is, $e^{i\phi}g$ is in the little group of 
$\d _{j} F ( \eta _0 )$.  
$\Box$

This theorem is particularly useful when $g$ is a non-trivial rotation
(and $\phi \ne \pi$); 
\ie ~when $\det \ghat =+1$ and $\ghat$ is not the identity matrix.  
Then the fixed points of $g$ are on the axis of rotation.  In this case,
the theorem states that on an axis of rotation, $\d _j F$ points along
the axis: $\d _j F( \eta _0 ) \propto \etabar _0$.  If we restrict to a 
surface of constant $|\eta|^2 = r^2$, then we find
\be
\left. \frac{\d F}{\d \eta _j} ( \eta _0 ) \right| _{|\eta| =r} = 0
\ee
on an axis of rotation.  The axes of rotation are critical points of
$F$ in the angular variables.

We again consider $(1,\omega , \omega ^2)$ of $\Gamma _4^-$ of $O_h$
as an example.  This is a fixed point of $C_3$ about the $(1,1,1)$
axis up to a phase which is a third root of unity:
$\Chat _3 (1,\omega , \omega ^2) = \omega (1,\omega , \omega ^2)$.
According to the theorem, the gradient of the free energy at
$(1,\omega , \omega ^2)$ must be invariant under $\omega \Chat _3$;
hence, it is proportional to the dual vector $(1,\omega ^2, \omega )$.
This radial vector projects to 0 when $r$ is held fixed, so
$(1,\omega , \omega ^2)$ is a critical point of the free energy.

Note that perturbation theory is not used directly in the proof of
these theorems.  To the extent that we can say that the functions of
interest lie in a particular irrep, the theorems are non-perturbative.

Theorem \ref{th-crit} also has implications for solutions of the
gap equation.
Since it is an auxiliary equation
of motion for the free energy, the theorem
implies that the full gap equation is stable for
$\Delta$ (or more precisely, $\eta _m(\Gamma )$) pointing along 
any of the rotational symmetry axes of
the representation space.  Stability means that
the function
\be
-\frac{1}{\Delta _{\vec{k}}} 
\sum_{\vec{k}\p} \frac {W_{\vec{k},\vec{k}\p}}
 {2E_{\vec{k}\p}} \Delta_{\vec{k}\p} 
\tanh \left( \txthalf \beta E_{\vec{k}\p} \right) 
\ee
is invariant under $\GG$.  This reduces the gap equation to a 
one dimensional problem, only slightly less tractable that the
usual singlet BCS solution.

\section{Some Exact Results}
\label{app-exact}

In this Appendix we present an exact solution of the gap
equation at zero temperature for two of the cases studied in Section
\ref{sec-Pairing}.  These two gap functions, $\dhat = (0,0,1)$ 
and $\dhat = (1,i,0)$ of 
$\Gamma _4^-$ of $O_h$, are plotted in Figure \ref{fig-gap},
along with the $\Gamma _1^-$ gap function, which has not been 
calculated in closed form. 
\be
\frac{1}{\lambda _{(0,0,1)}} = 
  \arcsinh ( \nu  /\sqrt{3}) + 
  {\tst \frac{1}{6}} \nu  \sqrt{3 + \nu ^2} 
 + \frac{\nu ^3}{6\sqrt{3}} \log \left( \frac{\nu }{\sqrt{3} + 
  \sqrt{3+\nu ^2}} \right)
\ee
where $\nu = \omega _c / \Delta _{rms}$.
\be
\frac{1}{\lambda _{(1,i,0)}} = 
\frac{\nu ^2}{6} + 
\frac{\nu  ( 9-2 \nu ^2)}{6 \sqrt{6}} 
  \arctan \left( \sqrt{\frac{3}{2}} \, \nu ^{-1} \right)
+ \txthalf \log \left( 1 + {\tst \frac{2}{3}} \nu ^2 \right)
\ee
It is evident from the plot that the magnitude of the gap function
at low temperature decreases as the number of gap function nodes
increases, but that the effect is not dramatic.

\section{The Rotationally Invariant Free Energy}
\label{app-O3}

In the text we constructed and analyzed the free energy with the gap
function in various irreducible representations of the cubic,
hexagonal and tetragonal point groups.  The resulting phase
diagram is quite complicated because of the many couplings 
necessary to specify the pairing interaction.  Only a small
subset of these couplings corresponds to physical perturbations.  

The free energy simplifies greatly if we make the 
natural ansatz that the dominant channel for pairing comes from 
the interaction (\ref{kxkp})
\be
W_{\vec k,\vec k\p} = -|{\WW}|\frac{\vec k \cdot \vec k\p}{k_F^2/3},
\label{leadInt}
\ee
and that the Fermi surface is spherical.
This interaction has an $O(3)$ symmetry, a
higher symmetry than we have considered in Section \ref{sec-GL}.
Let the gap function be 
in $\Gamma _4^-$ of $O_h$, the only cubic irrep that
pairs under the interaction (\ref{leadInt}).  
It has two non-cyclic degrees of freedom, the magnitude, $r$, 
and the angle, $\varphi$, between its real and imaginary parts:
\be
\Delta = r \left( 1+i \cos \varphi , i \sin \varphi , 0 \right)\cdot \vec{k} 
/ \sqrt{2}.  
\ee
The overall phase and the three Euler angles do not affect the
energy.
The free energy is expressed in terms of the well-known $O(3)$ invariant
polynomials \cite{FHappF}
\be
\barcl
{\dst P_1} & = & 
{\dst |\eta _1|^{2} + |\eta _2|^{2} + |\eta _3|^{2} ~=~r^2} \\ [2mm]
{\dst P_2} & = & 
{\dst  |\eta _1^{2} + \eta _2^{2} + \eta _3^{2}|^2 
 ~=~ r^4 \cos ^2 \varphi } 
\eac
\ee
and it is given by
\be
\barcl
{\dst F} & = & {\dst
\alpha r^2 + \beta _1 r^4
 +\beta _2 r^4 \cos ^2 \varphi
+ \gamma _1 r^6 +
\gamma _2 r^6 \cos ^2 \varphi
+ \cdots
}
\eac
\ee

According to Theorem \ref{th-crit} (extended to $O(3)$), the 
generic critical points are $r_0(0,0,1)\cdot \vec{k}$ and
$r_0(1,i,0)\cdot \vec{k}/\sqrt{2}$ up to rotation.  
Note that these have a line of zeros and 2 point zeros, respectively.
The free energy has non-generic critical points in small regions 
of parameter space at eighth order in $\Delta$.  

At sixth order, the ground state gap function is given by
\be
\bal
{\dst \Delta = r_0 k_z  \hspace{2.8cm} \beta _{2R} ~\le ~ 0} \\ [2mm]
{\dst \Delta = r_0 (k_x + i k_y)/\sqrt{2} ~~~~~ \beta _{2R} ~\ge ~0} \\ [2mm]
 ~~~~~~ {\dst {\mathrm where} ~~ 
\beta _{2R} = -\left[ 4\Phi 
(\beta _1 + \beta _2 , |\alpha | ( \gamma _1 + \gamma _2 )/
(\beta _1 + \beta _2)^2)\right] ^{-1} }
\eac
\ee
and note that $\beta _{2R} = \beta _{2}$ at fourth order.
The value of $r_0$ is given by the expression in Eq.\ (\ref{r0six}) with
$\beta = \beta _1$ and $\gamma = \gamma _1$ in the first case, and
$\beta = \beta _1+\beta _2$ and $\gamma = \gamma _1+\gamma _2$ in the 
second.

\pagebreak

\begin{table}
\caption[]{Categorization of normal state symmetries, number of allowed
broken symmetry broken states, and allowed values of S and L for
several fermion liquids (S, L values for the BCS case refer to cubic
crystal symmetry).  ``HFS'' refers to the picture where the spin
is frozen into the lattice and therefore is not a separate symmetry of
the normal state; it is still unclear if this picture gives the best
description of the heavy fermion superconductors.  ``No $\cal{I}$''
indicates ``HFS'' without inversion (see text).
Symmetry group
notation is given in the text.\label{table-sym}} 

\begin{tabular}{cccc}
System & Normal & Broken Symmetries & Pairing Type \\
\hline
$^3$He & ${\cal L}\times{\cal S}\times{\cal T}\times{\cal I}\times$ U(1) 
       & $\infty$ & S=0, L=even  \\
       &     &       & S=1, L=odd   \\

 BCS   & ${\cal G}\times{\cal S}\times{\cal T}\times{\cal I}\times$ U(1) 
       &  Finite     & S=0, L=0,2,4,6  \\
       &     &       & S=1, L=1,3,5,9  \\

 HFS   & ${\cal G}\times{\cal T}\times{\cal I}\times$ U(1) 
       &  Fewer      & S,L Coupled \\
       & &  & Even or Odd Parity \\

 No~$\II$   & ${\cal G}\times{\cal T}\times$ U(1) 
       &  Still Fewer      & Impure States \\

 SSS   & ${\cal G}\times{\cal I}\times$ U(1) 
       &  Fewest     & L=odd  \\

\end{tabular}
\end{table}

\pagebreak
\begin{table}
\tightenlines
\caption[]{\label{table-basis}Leading basis functions for 
the irreps of $O_h$, $D_{6h}$ and $D_{4h}$.}
\asize{1.0}
\begin{tabular}{ll} 
Irrep $\Gamma$ & Basis $\psi ( \Gamma, m; \vec{k} )$ \\ \hline 
\multicolumn{2}{c}{Octagonal --- $O_{h}$} \\ \\
$\Gamma _1^-$ & 
$\psi ( \Gamma _1^-, 1; \vec{k} ) = 
k_xk_yk_z ( k_x^2 - k_y^2 )( k_y^2 - k_z^2 )( k_z^2 - k_x^2)$ \\ [2mm]
$\Gamma _2^-$ & $\psi ( \Gamma _2^-, 1; \vec{k} ) =k_xk_yk_z$ \\ [2mm]
$\Gamma _3^-$ & 
 $\psi ( \Gamma _3^-, 1; \vec{k} ) =k_xk_yk_z(2k_z^2 - k_x^2- k_y^2)$ \\ 
 & $\psi ( \Gamma _3^-, 2; \vec{k} ) = \sqrt{3} k_xk_yk_z(k_x^2 - k_y^2)$ \\
[2mm]
 &
 $\psi ( \Gamma _3^-, 1\p; \vec{k} ) =
k_xk_yk_z \left( k_z^2 + \omega k_x^2 + \omega ^2 k_y^2 \right)
 $ \\
 & $\psi ( \Gamma _3^-, 2\p; \vec{k} ) =
k_xk_yk_z (k_z^2 + \omega  ^2 k_x^2 + \omega k_y^2)
 $ \\
[2mm]
$\Gamma _4^-$ & $\psi ( \Gamma _4^-, 1; \vec{k} ) = k_x $ \\  
 & $\psi ( \Gamma _4^-, 2; \vec{k} ) = k_y $ \\ 
 & $\psi ( \Gamma _4^-, 3; \vec{k} ) = k_z $ \\  [2mm]
$\Gamma _5^-$ & $\psi ( \Gamma _5^-, 1; \vec{k} ) =k_x(k_y^2 - k_z^2) $ \\ 
 & $\psi ( \Gamma _5^-, 2; \vec{k} ) =k_y(k_z^2 - k_x^2) $ \\ 
 & $\psi ( \Gamma _5^-, 3; \vec{k} ) =k_z(k_x^2 - k_y^2) $ \\ \\ \hline
\multicolumn{2}{c}{Hexagonal --- $D_{6h}$} \\ \\ 
$\Gamma _1^-$ & 
$\psi ( \Gamma _1^-, 1; \vec{k} ) = 
k_xk_yk_z ( k_x^2 - 3k_y^2 )( k_y^2 - 3k_x^2 )$ \\ [2mm]
$\Gamma _2^-$ & $\psi ( \Gamma _2^-, 1; \vec{k} ) =k_z$ \\ [2mm]
$\Gamma _3^-$ & 
 $\psi ( \Gamma _3^-, 1; \vec{k} ) =k_y^3 - 3k_x^2 k_y$ \\ [2mm]
$\Gamma _4^-$ & 
 $\psi ( \Gamma _4^-, 1; \vec{k} ) =k_x^3 - 3k_x k_y^2$ \\ [2mm]
$\Gamma _5^-$ & $\psi ( \Gamma _5^-, 1; \vec{k} ) = k_x $ \\  
 & $\psi ( \Gamma _5^-, 2; \vec{k} ) = k_y $ \\  [2mm]
$\Gamma _6^-$ & $\psi ( \Gamma _6^-, 1; \vec{k} ) =
 k_xk_yk_z(k_y^2 - 3k_x^2) $ \\ 
 & $\psi ( \Gamma _6^-, 2; \vec{k} ) =
 k_y^2k_z(k_y^2 - 3k_x^2) $ \\ \\
\hline
\multicolumn{2}{c}{Tetragonal --- $D_{4h}$} \\ \\
$\Gamma _1^-$ & 
$\psi ( \Gamma _1^-, 1; \vec{k} ) = 
k_xk_yk_z ( k_x^2 - k_y^2 )$ \\ [2mm]
$\Gamma _2^-$ & $\psi ( \Gamma _2^-, 1; \vec{k} ) =k_z$ \\ [2mm]
$\Gamma _3^-$ & 
 $\psi ( \Gamma _3^-, 1; \vec{k} ) =k_xk_yk_z$ \\ [2mm]
$\Gamma _4^-$ & 
 $\psi ( \Gamma _4^-, 1; \vec{k} ) =(k_x^2 - k_y^2)k_z$ \\ [2mm]
$\Gamma _5^-$ & $\psi ( \Gamma _5^-, 1; \vec{k} ) = k_x $ \\  
 & $\psi ( \Gamma _5^-, 2; \vec{k} ) = k_y $ \\  \\
\end{tabular}
\end{table}

\pagebreak
\begin{table}
\caption[]{\label{table-phases}SSS symmetry breaking phases.
} 
\asize{1.0}
\begin{tabular}{llllll} 
$\Gamma$ & $H (\Gamma ^{\prime} )$ &
$H_{phys}$ & $\beta _i$ &\#& $\Psi(\vec{k})$ \\ \hline
\multicolumn{6}{c}{Octahedral -- $O_h$} \\ \\
$\Gamma _1^-$ & $O_h( \Gamma _1^- )$ & $O_h$ & --- & 1 &
$k_xk_yk_z ( k_x^2 - k_y^2 )( k_y^2 - k_z^2 )( k_z^2 - k_x^2)$ \\ [2mm]
$\Gamma _2^-$ & $O_h( \Gamma _2^- )$ & $O_h$ & --- & 1 &
$k_xk_yk_z$ \\ [2mm]
$\Gamma _3^-$ & $D_{4h}( \Gamma _1^- )$ & $D_{4h}$ 
& $\beta_{2R} <0$,$\gamma _{3R} >0$ & 3 & $\sqrt{3}k_xk_yk_z(k_x^2 - k_y^2)$ \\ 
 & $D_{4h}( \Gamma _3^- )$ & $D_{4h}$ 
& $\beta_{2R} <0$,$\gamma _{3R} <0$ & 3 & $k_xk_yk_z(2k_z^2 - k_x^2- k_y^2)$ \\ 
 & $T_{h}( \Gamma _2^- )$ & $O_{h}$ 
& $\beta _{2R} >0$ & 1 & $k_xk_yk_z(k_z^2 + \omega k_x^2 + \omega ^2 k_y^2)$ \\ 
 & $T_{h}( \Gamma _3^- )$ & $O_{h}$ 
&  & 1 & $k_xk_yk_z(k_z^2 + \omega ^2 k_x^2 + \omega k_y^2)$ \\ [2mm]
$\Gamma _4^-$ & $C_{3i}( \Gamma _2^- )$ & $D_{3d}$ 
& $\beta_{3R} < 0 <\beta_{2R}$ & 4 & $k_z + \omega k_x + \omega ^2 k_y$ \\ 
 & $C_{3i}( \Gamma _3^- )$ & $D_{3d}$ 
&  & 4 & $k_z + \omega ^2 k_x + \omega k_y$ \\ 
 & $D_{3d}( \Gamma _2^- )$ & $D_{3d}$ 
& $\beta_{2R}, \beta_{3R} < 0 $ & 4 & $k_z + k_x + k_y$ \\ 
 & $D_{4h}( \Gamma _2^- )$ & $D_{4h}$ 
& $4\beta_{2R} < \beta_{3R}, \beta_{3R} > 0 $ & 3 & $k_z $ \\ 
 & $C_{4h}( \Gamma _3^- )$ & $D_{4h}$ 
& $ 0 < \beta_{3R} < 4\beta_{2R} $ & 3 & $k_x + i k_y $ \\ 
 & $C_{4h}( \Gamma _4^- )$ & $D_{4h}$ 
&  & 3 & $k_x - i k_y $ \\  
 & $D_{2h}( \Gamma _2^- )$ & $D_{2h}$ 
& $ \beta_{3R}=0,\beta _{2R} <0$ & 3 & $k_x + k_y $ \\ 
 & $D_{2h}( \Gamma _4^- )$ & $D_{2h}$ 
&  & 3 & $k_x - k_y $ \\  [2mm]
$\Gamma _5^-$ & $C_{3i}( \Gamma _2^- )$ & $D_{3d}$ 
& $\beta_{3R} < 0 <\beta_{2R}$ 
& 4 & $k_z(k_x^2 - k_y^2) + \omega k_x (k_y^2 - k_z^2) + 
 \omega ^2 k_y (k_z^2 - k_x^2)$ \\ 
 & $C_{3i}( \Gamma _3^- )$ & $D_{3d}$ 
&  & 4 & $k_z(k_x^2 - k_y^2) + \omega ^2 k_x (k_y^2 - k_z^2) + 
 \omega k_y (k_z^2 - k_x^2)$ \\ 
 & $D_{3d}( \Gamma _1^- )$ & $D_{3d}$ 
& $\beta_{2R}, \beta_{3R} < 0 $  & 4 & $k_z(k_x^2 - k_y^2) 
+ k_x (k_y^2 - k_z^2) + k_y (k_z^2 - k_x^2)$ \\ 
 & $D_{4h}( \Gamma _4^- )$ & $D_{4h}$ 
& $4\beta_{2R} < \beta_{3R}, \beta_{3R} > 0 $  & 3 & $k_z(k_x^2 - k_y^2) $ \\ 
 & $C_{4h}( \Gamma _4^- )$ & $D_{4h}$
& $ 0 < \beta_{3R} < 4\beta_{2R} $ & 3 & 
 $k_x (k_y^2 - k_z^2) + i k_y (k_z^2 - k_x^2)$ \\
 & $C_{4h}( \Gamma _3^- )$ & $D_{4h}$
&  & 3 & $k_x (k_y^2 - k_z^2) - i k_y (k_z^2 - k_x^2)$ \\ 
 & $D_{2h}( \Gamma _2^- )$ & $D_{2h}$
& $ \beta_{3R}=0,\beta _{2R}<0 $ & 3 & 
 $k_x (k_y^2 - k_z^2) + k_y (k_z^2 - k_x^2)$ \\
 & $D_{2h}( \Gamma _4^- )$ & $D_{2h}$
&  & 3 & $k_x (k_y^2 - k_z^2) - k_y (k_z^2 - k_x^2)$ \\ \\
\multicolumn{6}{c}{$\omega = e^{2\pi i/3}$} 
\end{tabular}
\end{table}

\pagebreak
\begin{table}
\tightenlines
\caption[]{\label{table-morePhases}SSS symmetry breaking phases (cont.).
} 
\asize{1.0}
\begin{tabular}{llllll} 
\multicolumn{6}{c}{Hexagonal --- $D_{6h}$} \\ \\
$\Gamma _1^-$ & $D_{6h}( \Gamma _1^- )$ & $D_{6h}$ & --- &
 1 & $k_xk_yk_z ( k_x^2 - 3k_y^2 )( k_y^2 - 3k_x^2 )$ \\ [2mm]
$\Gamma _2^-$ & $D_{6h}( \Gamma _2^- )$ & $D_{6h}$ & --- &
 1 & $k_z$ \\ [2mm]
$\Gamma _3^-$ & $D_{6h}( \Gamma _3^- )$ & $D_{6h}$ & --- &
 1 & $k_y^3 - 3 k_x^2k_y$ \\ [2mm]
$\Gamma _4^-$ & $D_{6h}( \Gamma _4^- )$ & $D_{6h}$ & --- &
 1 & $k_x^3 - 3 k_xk_y^2$ \\ [2mm]
$\Gamma _5^-$ & $D_{2h}( \Gamma _4^- )$ & $D_{2h}$ 
& $\beta _{2R} <0, \gamma _{3R}<0$ & 3 & $k_x$ \\ 
 & $D_{2h}( \Gamma _2^- )$ & $D_{2h}$ 
& $\beta _{2R} <0, \gamma _{3R}>0$ & 3 & $k_y$ \\ 
 & $C_{6h}( \Gamma _5^- )$ & $D_{6h}$ 
& $\beta _{2R} > 0$ & 1 & $k_x + ik_y$ \\ 
 & $C_{6h}( \Gamma _6^- )$ & $D_{6h}$ 
&  & 1 & $k_x - ik_y$ \\ [2mm]
$\Gamma _6^-$ & $D_{2h}( \Gamma _1^- )$ & $D_{2h}$ 
& $\beta _{2R} <0, \gamma _{3R}<0$ & 3 & $k_xk_yk_z ( k_y^2 - 3k_x^2 )$ \\ 
 & $D_{2h}( \Gamma _3^- )$ & $D_{2h}$ 
& $\beta _{2R} <0, \gamma _{3R}>0$ & 3 & $k_y^2k_z ( k_y^2 - 3k_x^2 )$ \\ 
 & $C_{6h}( \Gamma _2^- )$ & $D_{6h}$ 
& $\beta _{2R} > 0$ & 1 & $k_yk_z (k_x + ik_y)( k_y^2 - 3k_x^2 )$ \\ 
 & $C_{6h}( \Gamma _3^- )$ & $D_{6h}$ 
&  & 1 & $k_yk_z (k_x - ik_y)( k_y^2 - 3k_x^2 )$ \\ \\ \hline
\multicolumn{6}{c}{Tetragonal --- $D_{4h}$} \\ \\
$\Gamma _1^-$ & $D_{4h}( \Gamma _1^- )$ & $D_{4h}$ & --- &
 1 &$k_xk_yk_z ( k_x^2 - k_y^2 )$ \\ [2mm]
$\Gamma _2^-$ & $D_{4h}( \Gamma _2^- )$ & $D_{4h}$ & --- &
 1 &$k_z$ \\ [2mm]
$\Gamma _3^-$ & $D_{4h}( \Gamma _3^- )$ & $D_{4h}$ & --- &
 1 &$k_xk_yk_z$ \\ [2mm]
$\Gamma _4^-$ & $D_{4h}( \Gamma _4^- )$ & $D_{4h}$ & --- &
 1 &$(k_x^2 - k_y^2)k_z$ \\ [2mm]
$\Gamma _5^-$ & $C_{4h}( \Gamma _3^- )$ & $D_{4h}$ 
& $\beta _{2R} > \beta _{3R}, \beta _{2R}>0$ &  1 &$k_x +ik_y$ \\ 
 & $C_{4h}( \Gamma _4^- )$ & $D_{4h}$ 
&  & 1 & $k_x -ik_y$ \\ 
 & $D_{2h}( \Gamma _2^- )$ & $D_{2h}$ 
& $\beta _{2R}, \beta _{3R} < 0$ & 1 & $k_x +k_y$ \\ 
 & $D_{2h}( \Gamma _4^- )$ & $D_{2h}$ 
&  & 1 & $k_x -k_y$ \\ 
 & $D_{2h}( \Gamma _2^- )$ & $D_{2h}$ 
& $\beta _{3R} > \beta _{2R}, \beta _{3R} >0$  & 1 & $k_y$ \\ 
 & $D_{2h}( \Gamma _4^- )$ & $D_{2h}$ 
&  & 1 & $k_x$  \\ 
\end{tabular}
\end{table}

\pagebreak
\begin{table}
\caption[]{\label{table-zero}Zeros of the gap function guaranteed by symmetry.} 
\asize{1.0}
\begin{tabular}{llll} 
$\Gamma$ & $H (\Gamma ^{\prime} )$ & Non-trivial Mappings & 
Generic Zeros \\ \hline
\multicolumn{4}{c}{Octahedral -- $O_h$} \\ \\
$\Gamma _1^-$ & $O_h( \Gamma _1^- )$ & 
$I,8S_6,3\sigma _h, S_4, 6\sigma _d$ &
9 circles: $\{ k_i = 0 \}_{i=1,2,3}, \{ k_i = \pm k_j \} _{i\ne j}$
 \\ [2mm]
$\Gamma _2^-$ & $O_h( \Gamma _2^- )$ &
$6C_4,6C_2^{\prime},I,8S_6,3\sigma _h$ &
3 circles: $\{ k_i = 0 \}_{i=1,2,3}$
 \\ [2mm]
$\Gamma _3^-$ & $D_{4h}( \Gamma _1^- )$ & 
$I,2S_4,\sigma _h,2\sigma _v,2\sigma _d$ & 
5 circles: $\{ k_i = 0 \}_{i=1,2,3}, 
\{ k_1 = \pm k_2 \}$ 
 \\ 
 & $D_{4h}( \Gamma _3^- )$ & 
$2C_4,2C_2^{\prime \prime},I,\sigma _h,2\sigma _v$ & 
3 circles and 8 points: $\{ k_i = 0 \}_{i=1,2,3}, 
\{ (\pm 1, \pm 1, \pm 1 )/\sqrt{3} \}$
 \\
 & $T_{h}( \Gamma _2^- )$ &  
$4C_3,4C_3^{-1},I,3\sigma _h,4S_6,4S_6^{-1}$ & 
3 circles and 8 points: $\{ k_i = 0 \}_{i=1,2,3}, 
\{ (\pm 1, \pm 1, \pm 1 )/\sqrt{3} \}$
 \\ 
 & $T_{h}( \Gamma _3^- )$ & 
$4C_3,4C_3^{-1},I,3\sigma _h,4S_6,4S_6^{-1}$ & 
3 circles and 8 points: $\{ k_i = 0 \}_{i=1,2,3}, 
\{ (\pm 1, \pm 1, \pm 1 )/\sqrt{3} \}$
 \\ [2mm]
$\Gamma _4^-$ & $C_{3i}( \Gamma _2^- )$ &  
$C_3,C_3^{-1},I,S_6,S_6^{-1}$ & 
2 points: $\{ \pm (1,1,1)/\sqrt{3} \}$
 \\
 & $C_{3i}( \Gamma _3^- )$ &  
$C_3,C_3^{-1},I,S_6,S_6^{-1}$ &
2 points: $\{ \pm (1,1,1)/\sqrt{3} \}$
 \\ 
 & $D_{3d}( \Gamma _2^- )$ & 
$3C_2^{\prime},I,2S_6 (\sigma _h~{\mathrm{of}}~D_{6d})$ & 
1 circle: $\{ k_x + k_y + k_z = 0 \}$ 
 \\
 & $D_{4h}( \Gamma _2^- )$ & 
$2C_2^{\prime},2C_2^{\prime \prime},I,2S_4,\sigma _h$ & 
1 circle: $\{ k_z = 0 \}$ 
 \\
 & $C_{4h}( \Gamma _3^- )$ & 
$C_4,C_2,C_4^{-1},I,S_4,S_4^{-1}$ & 
2 points: $\{ \pm (0,0,1) \}$
 \\
 & $C_{4h}( \Gamma _4^- )$ & 
$C_4,C_2,C_4^{-1},I,S_4,S_4^{-1}$ & 
2 points: $\{ \pm (0,0,1) \}$
 \\  [2mm]
$\Gamma _5^-$ & $C_{3i}( \Gamma _2^- )$ &  
$C_3,C_3^{-1},I,S_6,S_6^{-1}$ &
2+12 points: 
$\{ (\pm 1,\pm 1,\pm 1)/\sqrt{3} \}$,
$\{ (0,0,\pm 1) \} + cyc.$
 \\ 
 & $C_{3i}( \Gamma _3^- )$ & 
$C_3,C_3^{-1},I,S_6,S_6^{-1}$ &
2+12 points: 
$\{ (\pm 1,\pm 1,\pm 1)/\sqrt{3} \}$,
$\{ (0,0,\pm 1) \} + cyc.$
 \\
 & $D_{3d}( \Gamma _1^- )$ & 
$I,2S_6,3\sigma _d$ & 
3 circles: $\{ k_i = k_j \} _{i\ne j} $
 \\
 & $D_{4h}( \Gamma _4^- )$ & 
$2C_4,2C_2^{\prime},I,2S_4,\sigma _h,2\sigma _d$ & 
3 circles: $\{ k_z = 0 \}, \{ k_x = \pm k_y \}$
 \\
 & $C_{4h}( \Gamma _3^- )$ & 
$C_4,C_2,C_4^{-1},I,S_4,S_4^{-1}$ & 
2+12 points: 
$\{ (0,0,\pm 1) \} + cyc.$,
$\{ (\pm 1,\pm 1,\pm 1)/\sqrt{3} \}$
 \\
 & $C_{4h}( \Gamma _4^- )$ & 
$C_4,C_2,C_4^{-1},I,S_4,S_4^{-1}$ & 
2+12 points: 
$\{ (0,0,\pm 1) \} + cyc.$,
$\{ (\pm 1,\pm 1,\pm 1)/\sqrt{3} \}$
\end{tabular}
\end{table}

\pagebreak
\begin{table}
\caption[]{\label{table-moreZero}Zeros of the gap function 
guaranteed by symmetry (cont.).} 
\begin{tabular}{llll} 
\multicolumn{4}{c}{Hexagonal --- $D_{6h}$} \\ \\
$\Gamma _1^-$ & $D_{6h}( \Gamma _1^- )$ & 
$I,\sigma _h,2S_3,2S_6,3\sigma _d,3\sigma _v$ &
7 circles: $\{k_i = 0 \}_{i=1,2,3}, \{ \sqrt{3} k_x = \pm k_y \},
\{ \sqrt{3} k_y = \pm k_x \}$
\\ [2mm]
$\Gamma _2^-$ & $D_{6h}( \Gamma _2^- )$ & 
$3C_2^{\prime},3C_2^{\prime \prime},I,\sigma _h,2S_3,2S_6$ &
1 circle: $\{k_z = 0 \}$
 \\ [2mm]
$\Gamma _3^-$ & $D_{6h}( \Gamma _3^- )$ & 
$C_2,2C_6,3C_2^{\prime \prime},I,2S_6,3\sigma _d$ &
3 circles: $\{k_y = 0 \}, \{ \sqrt{3} k_x = \pm k_y \}$
 \\ [2mm]
$\Gamma _4^-$ & $D_{6h}( \Gamma _4^- )$ & 
$C_2,2C_6,3C_2^{\prime},I,2S_6,3\sigma _v$ &
3 circles: $\{k_x = 0 \}, \{ \sqrt{3} k_y = \pm k_x \}$
 \\ [2mm]
$\Gamma _5^-$ & $D_{2h}( \Gamma _2^- )$ & 
$C_2,C_2^{\prime \prime},I,\sigma _v^{\prime}$ &
1 circle: $\{k_y = 0 \}$
 \\ 
 & $D_{2h}( \Gamma _4^- )$ & 
$C_2,C_2^{\prime},I,\sigma _v^{\prime \prime}$ &
1 circle: $\{k_x = 0 \}$
 \\ 
 & $C_{6h}( \Gamma _5^- )$ & 
$C_2,C_3,C_3^{-1},C_6,C_6^{-1},I,S_3,S_3^{-1},S_6,S_6^{-1}$ &
2 points: $\{ \pm (0,0,1) \}$
 \\ 
 & $C_{6h}( \Gamma _6^- )$ & 
$C_2,C_3,C_3^{-1},C_6,C_6^{-1},I,S_3,S_3^{-1},S_6,S_6^{-1}$ &
2 points: $\{ \pm (0,0,1) \}$
 \\ [2mm]
$\Gamma _6^-$ & $D_{2h}( \Gamma _1^- )$ & 
$I,\sigma _v,\sigma _v^{\prime},\sigma _v^{\prime \prime}
(\sigma _d^{\prime},\sigma _d^{\prime \prime}~{\mathrm{of}}~ D_{6h})$ &
3+2 circles: $\{k_i = 0 \}_{i=1,2,3}$ + $\{k_y = \pm \sqrt{3} k_x \}$
\\ 
 & $D_{2h}( \Gamma _3^- )$ & 
$C_2^{\prime},C_2^{\prime \prime},I,\sigma _v
(\sigma _d,\sigma _d^{\prime},\sigma _d^{\prime \prime}~{\mathrm{of}}~D_{6h})$ &
1+3 circles: $\{k_3 = 0 \}$ +
$\{k_2 = 0 \}$, $\{k_y = \pm \sqrt{3} k_x \}$
 \\ 
 & $C_{6h}( \Gamma _2^- )$ & 
$C_3,C_3^{-1},C_6,C_6^{-1},I,S_3,S_3^{-1},S_6,S_6^{-1},\sigma _h$ &
1+3 circles: $\{k_3 = 0 \}$ +
$\{k_2 = 0 \}$, $\{k_y = \pm \sqrt{3} k_x \}$
 \\ 
 & $C_{6h}( \Gamma _3^- )$ & 
$C_3,C_3^{-1},C_6,C_6^{-1},I,S_3,S_3^{-1},S_6,S_6^{-1},\sigma _h$ &
1+3 circles: $\{k_3 = 0 \}$ +
$\{k_2 = 0 \}$, $\{k_y = \pm \sqrt{3} k_x \}$
\\ \\ \hline
\multicolumn{4}{c}{Tetragonal --- $D_{4h}$} \\ \\
$\Gamma _1^-$ & $D_{4h}( \Gamma _1^- )$ &
$I,2S_4,\sigma _h,2\sigma _v,2\sigma _d$ &
5 circles: $\{ k_i = 0 \}_{i=1,2,3}, \{ k_x = \pm k_y \}$
 \\ [2mm]
$\Gamma _2^-$ & $D_{4h}( \Gamma _2^- )$ & 
$2C_2^{\prime},2C_2^{\prime \prime},I,2S_4,\sigma _h$ &
1 circle: $\{ k_z = 0 \}$ 
 \\ [2mm]
$\Gamma _3^-$ & $D_{4h}( \Gamma _3^- )$ & 
$2C_4,2C_2^{\prime \prime},I,\sigma _h,2\sigma _v$ & 
3 circles: $\{ k_i = 0 \}_{i=1,2,3}$ 
 \\ [2mm]
$\Gamma _4^-$ & $D_{4h}( \Gamma _4^- )$ & 
$2C_4,2C_2^{\prime},I,2S_4,\sigma _h,2\sigma _d$ &
3 circles: $\{ k_z = 0 \}, \{ k_x = \pm k_y \}$
 \\ [2mm]
$\Gamma _5^-$ & $C_{4h}( \Gamma _3^- )$ & 
$C_4,C_2,C_4^{-1},I,S_4,S_4^{-1}$ &
2 points: $\{ \pm (0,0,1) \}$
 \\
 & $C_{4h}( \Gamma _4^- )$ &
$C_4,C_2,C_4^{-1},I,S_4,S_4^{-1}$ &
2 points: $\{ \pm (0,0,1) \}$
 \\
 & $D_{2h}( \Gamma _2^- )$ &
$C_2,C_2^{\prime \prime},I,\sigma _d^{\prime}$ &
1 circle: $\{ k_x = -k_y \}$ 
 \\
 & $D_{2h}( \Gamma _4^- )$ & 
$C_2,C_2^{\prime},I,\sigma _d^{\prime \prime}$ &
1 circle: $\{ k_x = k_y \}$ 
 \\
 & $D_{2h}( \Gamma _2^- )$ & 
$C_2,C_2^{\prime \prime},I,\sigma _v^{\prime}$ &
1 circle: $\{ k_y = 0 \}$ 
 \\
 & $D_{2h}( \Gamma _4^- )$ & 
$C_2,C_2^{\prime},I,\sigma _v^{\prime \prime}$ &
1 circle: $\{ k_x = 0 \}$ 
 \\ 
\end{tabular}
\end{table}

\pagebreak

\pagebreak

\begin{figure}[tbp]
\epsfysize=15cm\centerline{\epsffile{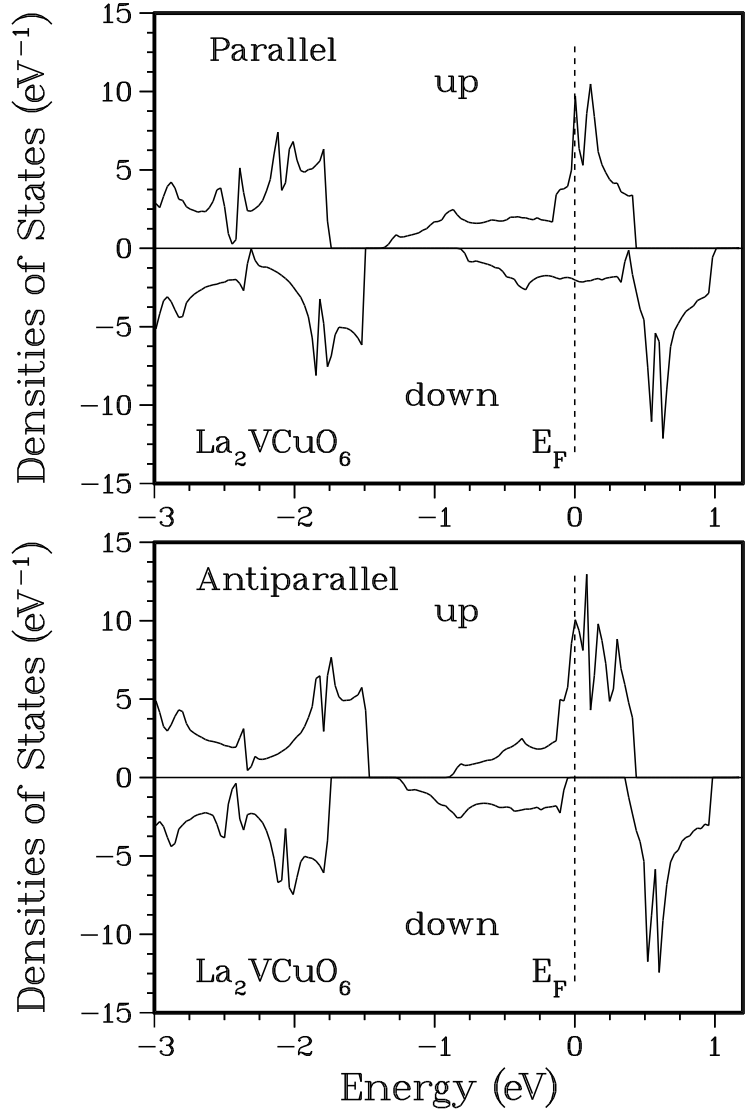}}
\caption{Total densities of states for each spin direction in the
double perovskite compound La$_2$VCuO$_6$.  Top panel: parallel alignment
of the Cu$^{2+}$ and V$^{4+}$ spins, with exchange splitting of
0.25 eV for the lower lying (Cu) states and 0.5 eV for the higher lying
(V) states.  Bottom panel: antiparallel alignment of the spins,
resulting in a HM AFM system with the Fermi level $E_F$ lying in the gap
of the down spin channel.
\label{fig-calcDOS}}
\end{figure}

\begin{figure}[tbp]
\epsfysize=15cm\centerline{\epsffile{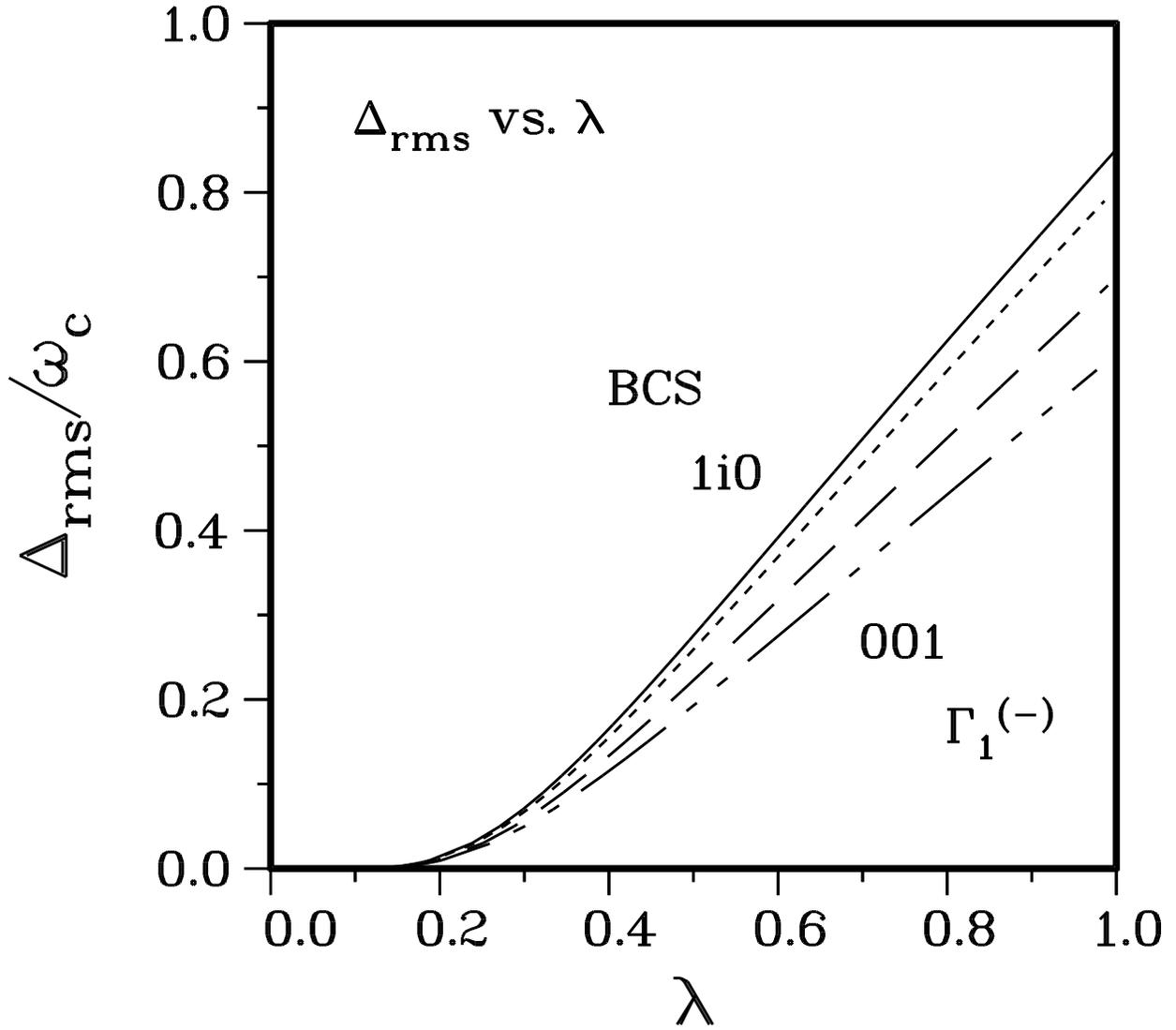}}
\caption{The RMS value of the gap function at $T=0$ relative to the 
coupling boson frequency cutoff, $\omega _c$, for the singlet
BCS case, and for three SSS cases.  The coupling strength $\lambda$ is
defined in the text.  Note that the point nodes of the ``1i0'' gap,
the line nodes of the ``001'' gap, and the set of {\em nine} line
nodes for the $\Gamma_1^{(-)}$ case do not affect the RMS gap value
drastically. 
\label{fig-gap}}
\end{figure}

\begin{figure}[tbp]
\epsfysize=15cm\centerline{\epsffile{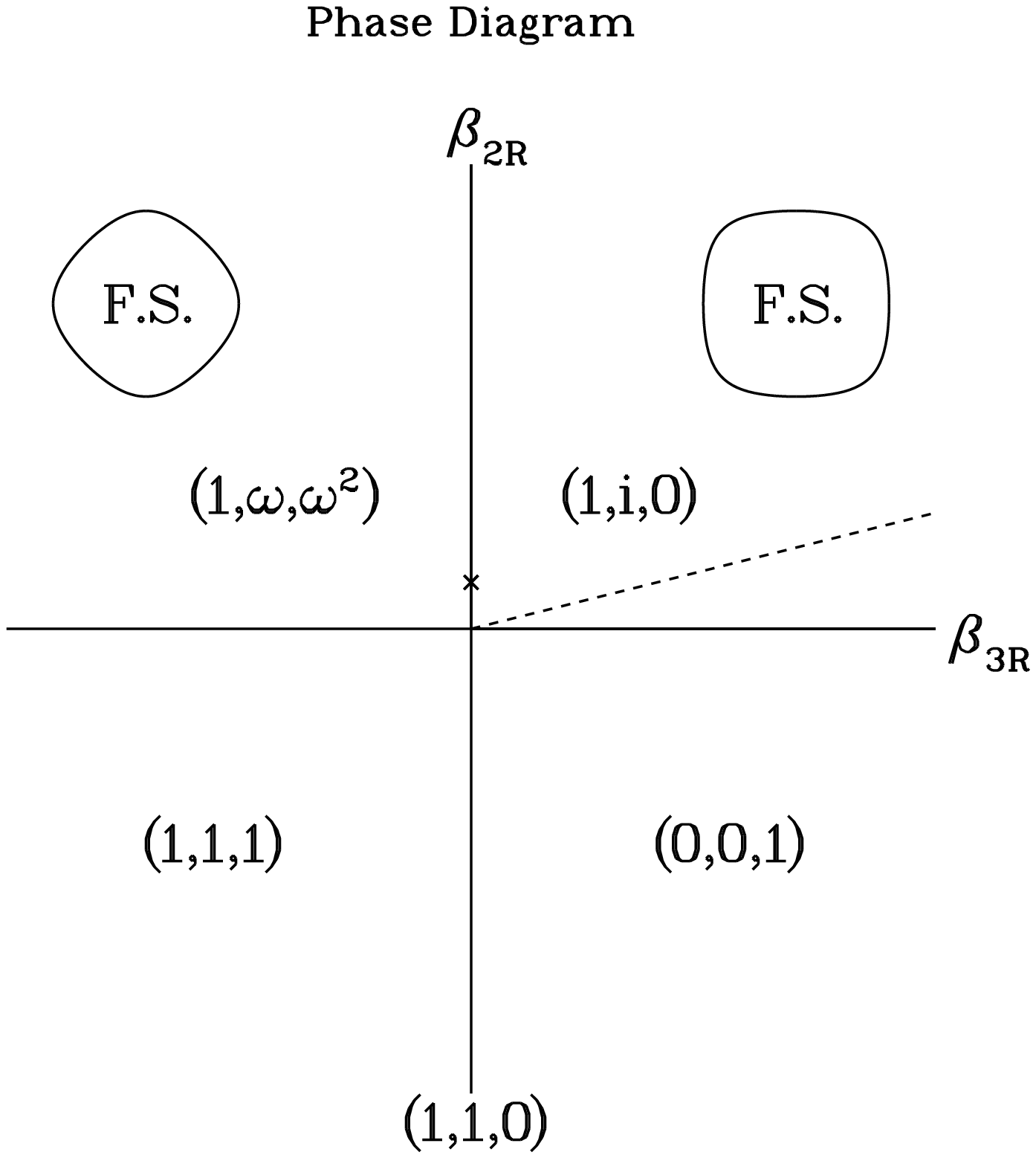}}
\caption{The Ginzburg-Landau phase diagram for SSS with $\Gamma _{4,5}^-$
cubic symmetry has five phases near $T_c$.  The cubic group is only
partly broken in each of these phases.  The notation is explained fully in
the text, but for $\Gamma _4^-$ the gap function is proportional to
$\hat{d}\cdot \vec{k}$, where $\hat{d}$ is the vector labelling each 
phase.  The phases at the top of the diagram have point nodes and
those at the bottom have line nodes.  The phase with $\hat{d}=(1,1,0)$
is only stable on the negative $\beta _{2R}$ axis (see the text).
The ``$\times$'' on the positive  $\beta _{2R}$ axis denotes the weak
coupling point with a spherical Fermi surface and the interaction
(\ref{FSHint}).  When the Fermi surface is deformed outward or inward
at the diagonals as pictured,
the ground state is a phase on the right or left side of the diagram,
respectively.
\label{fig-cubPhase}}
\end{figure}


\begin{thebibliography}{99}
\bibitem{foot-equivar}  Even when $\Gamma$ is one 
dimensional, $\Delta$  may pick up a phase under point group
transformations.  To be precise, $\ghat : \Delta \ra \rho _g\Delta$
where $\rho _g$ is a root of unity.  In this case, $\Delta$ is said to be 
equivariant (invariant up to a gauge transformation), and the symmetry
breaking mixes the gauge and point groups.  The result is that the
ground state possesses a non-trivial topological structure.
\end{thebibliography}
\end{document}